\RequirePackage{fix-cm}
\documentclass[smallextended]{svjour3}       
\smartqed  
\usepackage{graphicx}

\usepackage[T1]{fontenc}
\usepackage[latin9]{inputenc}
\usepackage[a4paper]{geometry}
\usepackage{array}
\usepackage{rotating}
\usepackage{float}
\usepackage{url}
\usepackage{multirow}
\usepackage{amsmath}
\usepackage{graphicx}
\usepackage{natbib}
\usepackage{amsmath,bm}
\usepackage{amssymb}
\usepackage[table]{xcolor}

\newcommand{\cst}{\mathrm{const}}
\newcommand{\Esp}{\mathrm{E}}

\begin{document}

\title{The Stochastic Topic Block Model for the Clustering of Vertices
  in Networks with
Textual Edges
}

\titlerunning{The Stochastic Topic Block Model} 

\author{C. Bouveyron \and
        P. Latouche \and R. Zreik
}

\institute{C. Bouveyron \at
              Laboratoire MAP5, UMR CNRS 8145\\
              Universit\'e Paris Descartes \& Sorbonne Paris Cit\'{e} \\
              \email{charles.bouveyron@parisdescartes.fr}            \\
           \and
           P. Latouche \at
              Laboratoire SAMM, EA 4543 \\
              Universit\'e Paris 1 Panth\'eon-Sorbonne\\
              \and
           R. Zreik \at 
              Laboratoire SAMM, EA 4543 \\
              Universit\'e Paris 1 Panth\'eon-Sorbonne\\
              Laboratoire MAP5, UMR CNRS 8145\\
              Universit\'e Paris Descartes \& Sorbonne Paris Cit\'{e} 
}

\date{Published in Statistics and Computing. The final publication is available at Springer via http://dx.doi.org/10.1007/s11222-016-9713-7"}

\maketitle

\begin{abstract}
Due to the significant increase of communications between individuals via
social media (Facebook, Twitter, Linkedin) or electronic formats (email, web, e-publication) in the past two decades,
network analysis has become a unavoidable discipline. Many random
graph models have been proposed to extract information from networks
based on person-to-person links only, without taking into account
information on the contents. This paper introduces the
stochastic topic block model (STBM), a probabilistic model for networks with textual edges. We address here the
problem of discovering meaningful clusters of vertices that are coherent from both the network interactions and the text contents. A classification
variational expectation-maximization (C-VEM) algorithm is proposed
to perform inference. Simulated data sets are considered in order
to assess the proposed approach and to highlight its main features. Finally, we demonstrate the effectiveness
of  our   methodology  on   two  real-word   data  sets:   a  directed
communication network and a undirected co-authorship network. 
\end{abstract}

\section{Introduction}

\begin{figure}
\begin{centering}
\includegraphics[width=1\columnwidth]{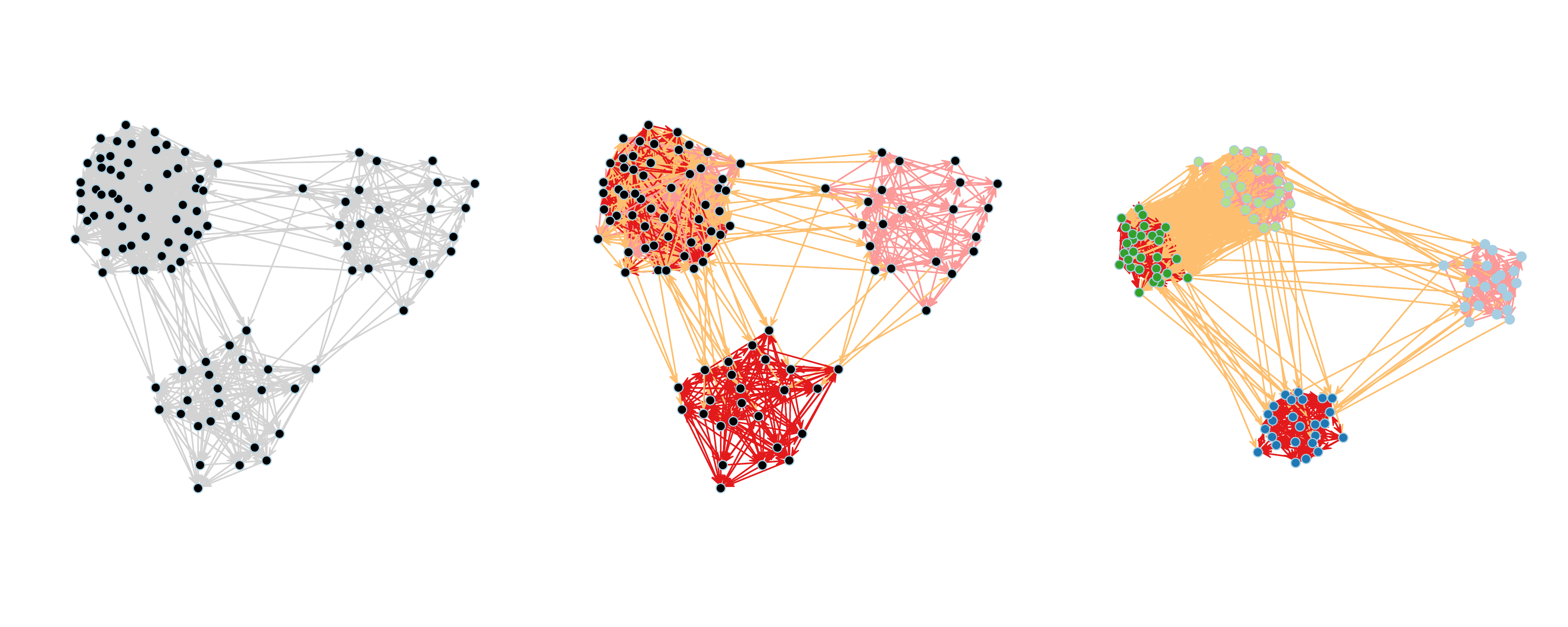}
\par\end{centering}

\protect\caption{\label{fig:A-sample-network}A sample network made of 3 ``communities''
where one of the communities is made of two topic-specific groups.
The left panel only shows the observed (binary) edges in the network.
The center panel shows the network with only the partition of edges
into 3 topics (edge colors indicate the majority topics of texts). The right
panel  shows the  network with  the clustering  of its  nodes (vertex colors indicate the groups) and  the
majority topic of the edges. The latter visualization allows
to see the topic-conditional structure of one of the three communities.}

\end{figure}

 The significant and recent increase of interactions between individuals via
social media or through electronic communications enables to observe frequently networks with textual edges. It is obviously
of strong  interest to be able to model  and cluster the  vertices of
those networks using information on both the network structure and the text contents.
Techniques able to provide such a clustering would allow a deeper
understanding of the studied networks. As a motivating example, Figure~\ref{fig:A-sample-network}
shows a network made of  three ``communities'' of vertices where one of the communities
can in fact be split into two separate groups based on the topics
of  communication  between  nodes  of  these  groups  (see  legend  of
Figure~\ref{fig:A-sample-network} for details). Despite the important
efforts in both network analysis and text analytics, only a few works
have focused on the joint modeling of network vertices
and textual edges.

\subsection{Statistical models for network analysis}

On the one hand, there is a long history of research in the statistical
analysis of networks,  which has received strong interest  in the last
decade. In particular,
statistical methods have imposed theirselves as efficient and flexible
techniques for network clustering. Most of those methods look for
specific structures, so called communities, which exhibit a transitivity
property such that nodes of the same community are more likely to
be connected~\citep{hofman2008bayesian}. Popular approaches for
community          discovering,         though          asymptotically
biased~\citep{bickel2009nonparametric}, are based on the modularity score given by~\citet{girvan2002community}.
Alternative clustering methods usually rely on the latent position cluster model
(LPCM) of~\citet{handcock2007model}, 
or the stochastic block model (SBM)~\citep{wang1987,nowicki2001estimation}.
The LPCM model, which extends the work of \cite{hoff2002latent}, assumes that the links between the vertices depend
on their positions in a social latent space and allows the simultaneous
visualization and clustering of a network. 

The SBM model is a flexible
random graph model which  is based on a probabilistic generalization of the method
applied by~\citet{white1976social} on Sampson's famous monastery~\citep{Samp}.
It assumes that each vertex belongs to a latent group, and that the
probability of connection between a pair of vertices depends exclusively
on their group. Because no specific assumption is made on the connection
probabilities, various types of structures of vertices can be taken
into account. At this point, it is important to notice that, in
  network clustering, two types of clusters are usually considered: communities (vertices within a community are more likely to
connect than vertices of different communities) and stars or disassortative clusters (the vertices
of a cluster highly connect to vertices of another). In this context, SBM is particularly useful in practice since it has the ability to characterize both types of clusters.

While SBM was originally developed to analyze mainly
binary networks, many extensions have been proposed since to deal
for instance with valued edges~\citep{articlemariadassou2010}, categorical
edges~\citep{RSM} or to take into account prior information~\citep{articlezanghi2010,matias2014}.
Note that other extensions of SBM have focused on looking for overlapping
clusters~\citep{airoldi2008mixed,latouche2009overlapping} or on
the modeling of dynamic networks~\citep{yang2011detecting,xu2013dynamic,Zreik2016,Matias2016}.

The inference of SBM-like models is usually done using variational
expectation maximization (VEM)~\citep{daudin2008mixture}, variational
Bayes EM (VBEM)~\citep{articlelatouche2012}, or Gibbs sampling~\citep{nowicki2001estimation}.
Moreover, we emphasize that various strategies have been derived to
estimates the number of corresponding clusters using model selection
criteria~\citep{daudin2008mixture,articlelatouche2012}, allocation
sampler~\citep{mcdaid11}, greedy search~\citep{come15}, or non
parametric schemes~\citep{kemp2006learning}. We refer to~\citep{salter2012}
for a overview of statistical models for network analysis.

\subsection{Statistical models for text analytics}

On the other hand, the statistical modeling of texts appeared at the
end of the last century with an early model described by~\citet{Papadimitriou}
for latent semantic indexing (LSI)~\citep{deerwester1990indexing}.
LSI is known in particular for allowing to recover linguistic notions
such as synonymy and polysemy from ``term frequency - inverse document
frequency'' (tf-idf) data. \citet{hofmann1999probabilistic} proposed
an alternative model for LSI, called probabilistic latent semantic
analysis (pLSI), which models each word within a document using a
mixture model. In pLSI, each mixture component is modeled by a multinomial
random variable and the latent groups can be viewed as \textquotedblleft topics\textquotedblright .
Thus, each word is generated from a single topic and different words
in a document can be generated from different topics. However, pLSI has no model at the document level and
may suffer from overfitting. Notice that pLSI can also be
viewed has an extension of the mixture of unigrams, proposed by~\citet{nigam2000text}.

The model which finally concentrates most desired features was proposed
by~\citet{LDA} and is called latent Dirichlet allocation (LDA).
The LDA model has rapidly become a standard tool in statistical text
analytics and is even used in different scientific fields such has
image analysis~\citep{lazebnik2006beyond} or transportation research~\citep{ComeLDA}
for instance. The idea of LDA is that documents are represented as
random mixtures over latent topics, where each topic is characterized
by a distribution over words. LDA is therefore similar to pLSI except
that    the   topic    distribution   in    LDA   has    a   Dirichlet
distribution. Several  inference procedures have been  proposed in the
literature   ranging  from   VEM   \citep{LDA}   to  collapsed   VBEM
\citep{teh}. 

Note that
a limitation of LDA would be the inability to take into account possible
topic correlations. This is due to the use of the Dirichlet distribution
to model the variability among the topic proportions. To overcome
this limitation, the correlated topic model (CTM) was also
developed by\emph{~}\citet{blei2006correlated}. Similarly, the relational
topic model (RTM)~\citep{RTM} models the links between documents
as binary random variables conditioned on their contents, but ignoring
the community ties between the authors of these documents. Notice
that the ``itopic'' model~\citep{itopicmodel} extends RTM to weighted
networks. The reader may refer to~\citet{blei2012probabilistic}
for an overview on probabilistic topic models.

\subsection{Statistical models for the joint analysis of texts and networks}

Finally, a few recent works have focused on the joint modeling of
texts and networks. Those works are mainly motivated by the will of
analyzing social networks, such as Twitter or Facebook, or electronic
communication networks. Some of them are partially based on LDA: the
author-topic (AT)~\citep{AT1,AT2} and the author-recipient-topic
(ART) \citep{ART} models. The AT model extends LDA to include authorship
information whereas the ART model includes authorships and information
about the recipients. Though potentially powerful, these models do
not take into account the network structure (communities, stars, ...)
while the concept of community is very important in the context
of social networks, in the sense that a community is a group of users
sharing similar interests. 

Among the most advanced models for the
joint analysis of texts and networks, the first models which explicitly
take into account both text contents and network structure are the
community-user-topic (CUT) models proposed by~\citep{CUT}. Two models
are proposed: CUT1 and CUT2, which differ on the way they construct
the communities. Indeed, CUT1 determines the communities only based
on the network structure whereas CUT2 model the communities based
on the content information solely. The CUT models  therefore deal
each with only a part of the problem we are interested in. It is also worth noticing that the authors of these models rely
for inference on Gibbs sampling which may prohibit their use on large
networks. 

A second attempt was made by~\citet{CART} who extended
the ART model by introducing the community-author-recipient-topic
(CART) model. The CART model adds to the ART model that authors and
recipients belong to latent communities and allows CART to recover
groups of nodes that are homogenous both regarding the network structure
and the message contents. Notice that CART allows the nodes to be
part of multiple communities and each couple of actors to have a specific
topic. Thus, though extremely flexible, CART is also a highly parametrized
model. In addition, the recommended inference procedure based on Gibbs
sampling may also prohibit its application to large networks. 

More
recently, the topic-link LDA~\citep{liu2009} also performs topic
modeling and author community discovery in a unified framework. As
its name suggests, topic-link LDA extends LDA with a community layer
where the link between two documents (and consequently its authors)
depends on both topic proportions and author latent features through
a logistic transformation. However, whereas CART focuses only on directed
networks, topic-link LDA is only able to deal with undirected networks.
On the positive side, the authors derive a variational EM algorithm
for inference, allowing topic-link LDA to eventually be applied to
large networks. 

Finally, a family of 4 topic-user-community models
(TUCM) were proposed by~\citet{TUCM}. The TUCM models are designed
such that they can find topic-meaningful communities in networks with
different types of edges. This in particular relevant in social networks
such as Twitter where different types of interactions (followers,
tweet, re-tweet, ...) exist. Another specificity of the TUCM models
is that they allow both multiple community and topic memberships.
Inference is also done here through Gibbs sampling, implying a possible
scale limitation.

\subsection{Contributions and organization of the paper}

We propose here a new generative model for the clustering of networks
with textual edges, such as communication or co-authorship networks.
Conversely to existing works which have either too simple or highly-parametrized
models for the network structure, our model relies for the network
modeling on the SBM model which offers a sufficient flexibility with
a reasonable complexity. This model is  one of the few 
able to recover different topological structures such as communities, stars or disassortative
clusters \citep[see][for instance]{articlelatouche2012}. Regarding the topic modeling, our approach is based on the
LDA model, in which the topics are conditioned on the latent groups.
Thus, the proposed modeling will be able to exhibit node partitions
that are meaningful both regarding the network structure and the topics,
with a model of limited complexity, highly interpretable, and for
both directed and undirected networks. In addition, the proposed inference
procedure -- a classification-VEM algorithm -- allows the use of our
model on large-scale networks. 

The proposed model, named stochastic topic block model (STBM), is
introduced in Section~2. The model inference is discussed in Section~3
as well as model selection. Section~4 is devoted to numerical experiments
highlighting the main features of the proposed approach and proving
the validity of the inference procedure. Two applications to real-world
networks (the Enron email and the Nips co-authorship networks) are
presented in Section~5. Section~6 finally provides some concluding
remarks.

\section{The model}

This section presents  the notations used in the paper  and introduces the
STBM model.  The joint distributions of  the model to create edges and
the corresponding documents are also given.

\subsection{Context and notations}

A directed network  with $M$ vertices, described by  its $M \times
M$ adjacency
matrix $A$, is  considered. Thus, $A_{ij}=1$ if there is  an edge from
vertex $i$ to vertex $j$,  $0$ otherwise.  The network  is  assumed not  to have  any
self-loop and therefore $A_{ii}=0$ for all $i$. If an edge  from $i$
to  $j$ is  present,  then it  is characterized  by  a set  of
$D_{ij}$  documents,  denoted $W_{ij}=(W_{ij}^{d})_d$.  Each  document
$W_{ij}^d$     is     made     of    a     collection     of $N_{ij}^{d}$    words
$W_{ij}^d=(W_{ij}^{dn})_n$.  In the directed scenario considered,
$W_{ij}$ can model  for instance a set of emails or text
messages  sent  from actor  $i$  to  actor  $j$.  Note that  all  the
methodology  proposed  in  this  paper easily  extends  to  undirected
networks.     In     such      a     case,     $A_{ij}=A_{ji}$     and
$W_{ij}^{d}=W_{ji}^{d}$ for all $i$ and $j$. The set $W_{ij}^{d}$ of documents can then
model for example  books or scientific papers written by  both $i$ and
$j$. In the  following, we denote $W=(W_{ij})_{ij}$ the  set of all
documents exchanged, for all the edges present in the network.

Our goal  is to cluster  the vertices  into $Q$ latent  groups sharing
homogeneous connection  profiles, \emph{i.e.} find an  estimate of the
set  $Y=(Y_1,  \dots,  Y_M)$  of latent  variables  $Y_i$  such  that
$Y_{iq}=1$ if vertex $i$ belongs to cluster $q$, and $0$ otherwise.
Although in some
cases, discrete or continuous edges are taken into account, the literature on networks focuses on modeling the presence of edges
as binary variables. The
clustering task  then consists in  building groups of  vertices having
similar  trends to  connect to  others.  In  this paper,  the connection
profiles are both characterized by the presence
of edges  and the documents  between pairs of vertices.  Therefore, we
aim  at  uncovering  clusters  by integrating  these  two  sources  of
information. Two nodes in the same  cluster should have the same trend
to  connect to  others, and  when  connected, the  documents they  are
involved in should be made of words related to similar topics.

\subsection{Modeling the presence of edges}\label{ssection:presEdges}

In order to  model the presence of edges between  pairs of vertices, a
stochastic   block  model   \citep{wang1987,nowicki2001estimation}  is
considered.  Thus, the  vertices are  assumed  to be  spread into  $Q$
latent clusters such that $Y_{iq}=1$  if vertex $i$ belongs to cluster
$q$,  and $0$  otherwise.  In  practice,  the binary  vector $Y_i$  is
assumed to be drawn from a multinomial distribution 
\begin{equation*}
  Y_i \sim \mathcal{M}\left(1,\rho=( \rho_1, \dots, \rho_Q)\right),
\end{equation*}
where $\rho$ denotes the vector of class proportions.  By  construction,
$\sum_{q=1}^{Q}\rho_q=1$ and $\sum_{q=1}^{Q}Y_{iq}=1, \forall i$.

An edge from $i$ to $j$ is then sampled from a Bernoulli distribution,
depending on their respective
clusters
\begin{equation}\label{eq:Acond}
  A_{ij} | Y_{iq}Y_{jr}=1 \sim \mathcal{B}( \pi_{qr}).
\end{equation}
In words, if  $i$ is in cluster  $q$ and $j$ in $r$,  then $A_{ij}$ is
$1$ with probability $\pi_{qr}$. In the following, we denote $\pi$ the
$Q \times Q$ matrix of connection probabilities.  Note that in the
undirected case, $\pi$ is symmetric. 

All   vectors   $Y_{i}$   are   sampled   independently,   and   given
$Y=(Y_1,\dots, Y_M)$, all edges in $A$ are assumed to be independent. This
leads to the following joint distribution
\begin{equation*}
  p(A,Y|\rho,\pi) = p(A|Y,\pi)p(Y|\rho),
\end{equation*}
where
\begin{equation*}
  \begin{aligned}[b]
    p(A|Y,\pi) &= \prod_{i\neq j}^M p(A_{ij}|Y_i,Y_j,\pi) \\
    &= \prod_{i\neq j}^M \prod_{q,l}^{Q}p(A_{ij}|\pi_{qr})^{Y_{iq}Y_{jr}},
  \end{aligned}
\end{equation*}
and
\begin{equation*}
  \begin{aligned}
p(Y|\rho) &= \prod_{i=1}^M p(Y_i|\rho) \\
    &= \prod_{i=1}^M\prod_{q=1}^{Q}\rho_q^{Y_{iq}}.
  \end{aligned}
\end{equation*}

\subsection{Modeling the construction of documents}\label{ssection:construcDoc}

As mentioned  previously, if  an edge  is present  from vertex  $i$ to
vertex   $j$,   then   a  set   of  documents   $W_{ij}=(W_{ij}^d)_d$,
characterizing the oriented pair $(i, j)$, is assumed
to be given.  Thus, in a  generative perspective, the edges in $A$ are
first  sampled using  previous section.  Given $A$,  the documents  in
$W=(W_{ij})_{ij}$ are
then  constructed.  The  generative   process  we  consider  to  build
documents is strongly related to the latent Dirichlet allocation (LDA) model
of \cite{LDA}.  The link  between STBM  and LDA is  made clear  in the
following section. The STBM model relies on two concepts at the core of the SBM and LDA
models  respectively. On  the one  hand, a  generalization of  the SBM
model would assume that any kind of relationships
between  two  vertices  can  be explained  by  their  latent  clusters
only. In the LDA model on the other hand, the main assumption is that words in documents
are drawn from a mixture distribution over topics, each document $d$ having
its own vector of topic proportions $\theta_d$. The STBM model combines these two
concepts  to introduce  a  new generative  procedure  for documents  in
networks. 

Each pair  of clusters $(q, r)$  of vertices is first  associated to a
vector of  topic proportions  $\theta_{qr}=(\theta_{qrk})_{k}$ sampled
independently from a Dirichlet
distribution
\begin{equation*}
  \theta_{qr} \sim \mathrm{Dir}\left(\alpha=(\alpha_1,\dots,\alpha_K)\right),
\end{equation*}
such that $\sum_{k=1}^{K}\theta_{qrk}=1, \forall (q,r)$.  We denote 
$\theta=(\theta_{qr})_{qr}$ and  $\alpha=(\alpha_1,\dots,\alpha_K)$ the  parameter vector
  controlling  the  Dirichlet  distribution.   Note that  in  all  our
  experiments we set each component of $\alpha$ to
  $1$ in order to obtain a uniform distribution. Since $\alpha$ is fixed, it
  does not  appear in  the conditional  distributions provided  in the
  following. 
  The $n$th word $W_{ij}^{dn}$ of documents $d$ in $W_{ij}$ is then
associated to a latent topic  vector $Z_{ij}^{dn}$ assumed to be drawn
from a multinomial distribution, depending on the latent vectors $Y_i$
and $Y_j$
\begin{equation}\label{eq:Zcond}
  Z_{ij}^{dn}|\left\{Y_{iq}Y_{jr}A_{ij}=1, \theta\right\}\sim
  \mathcal{M}\left(1,\theta_{qr}=( \theta_{qr1}, \dots, \theta_{qrK})\right).
\end{equation}
Note that $\sum_{k=1}^{K}Z_{ij}^{dnk}=1, \forall (i,j,d), A_{ij}=1$.  Equations
(\ref{eq:Acond}) and (\ref{eq:Zcond}) are related: they both involve
the  construction  of  random   variables  depending  on  the  cluster
assignment  of vertices  $i$  and $j$.  Thus, if  an  edge is  present
($A_{ij}=1$) and if $i$ is in cluster $q$ and $j$ in $r$, then the word
$W_{ij}^{dn}$  is in  topic  $k$  ($Z_{ij}^{dnk}=1$) with  probability
$\theta_{qrk}$.

Then, given  $Z_{ij}^{dn}$, the  word $W_{ij}^{dn}$  is assumed  to be
drawn from a multinomial distribution
\begin{equation}\label{eq:Wcond}
  W_{ij}^{dn}|Z_{ij}^{dnk}=1 \sim \mathcal{M}\left(1, \beta_k=(\beta_{k1},\dots,\beta_{kV})\right),
\end{equation}
where  $V$  is the  number  of  (different)  words in  the  vocabulary
considered  and  $\sum_{v=1}^{V}\beta_{kv}=1,\forall  k$  as  well  as
$\sum_{v=1}^{V}W_{ij}^{dnv}=1, \forall (i,j,d,n)$.  Therefore, if $W_{ij}^{dn}$ is from topic $k$, then it is
associated  to  word $v$  of  the  vocabulary ($W_{ij}^{dnv}=1$)  with
probability     $\beta_{kv}$.    Equations     (\ref{eq:Zcond})    and
(\ref{eq:Wcond}) lead  to the following  mixture model for  words over
topics
\begin{equation*}
  W_{ij}^{dn}|\left\{Y_{iq}Y_{jr}A_{ij}=1, \theta\right\}\sim
  \sum_{k=1}^{K}\theta_{qrk}\mathcal{M}\left(1, \beta_k\right),
\end{equation*}
where the $K \times V$ matrix $\beta=(\beta_{kv})_{kv}$ of
probabilities does  not depend on  the cluster assignments.  Note that
words of different documents $d$ and $d^{'}$ in $W_{ij}$ have the same
mixture distribution which only depends  on the respective clusters of
$i$ and $j$. 
We also point out that words of the vocabulary appear in any document $d$ of $W_{ij}$ with
probabilities
\begin{equation*}
  \mathbb{P}(W_{ij}^{dnv}=1|Y_{iq}Y_{jr}A_{ij}=1,\theta )=\sum_{k=1}^{K}\theta_{qrk}\beta_{kv}.
\end{equation*}
Because  pairs $(q,  r)$ of  clusters  can have  different vectors  of
topics proportions  $\theta_{qr}$, the  documents they  are associated
with 
can have different mixture distribution of words over topics.  For instance,
most  words exchanged  from vertices  of  cluster $q$  to vertices  of
cluster $r$ can  be related to \emph{mathematics}  while vertices from
$q'$ can  discuss with  vertices of $r'$  with words  related to
\emph{cinema} and in some cases to \emph{sport}.

All the latent variables $Z_{ij}^{dn}$ are assumed to be sampled
independently and, given the latent variables, the words $W_{ij}^{dn}$
are  assumed   to  be  independent.  
Denoting $Z=(Z_{ij}^{dn})_{ijdn}$, this leads to the following joint distribution
\begin{equation*}
  p(W, Z, \theta|A, Y, \beta)=p(W|A, Z, \beta)p(Z|A, Y,\theta)p(\theta),
\end{equation*}
where
\begin{equation*}
  \begin{aligned}
  p(W|A, Z, \beta)&=\prod_{i \neq j}^{M}
  \left\{\prod_{d=1}^{D_{ij}}\prod_{n=1}^{N_{ij}^{d}}p(W_{ij}^{dn}|Z_{ij}^{dn},
    \beta)\right\}^{A_{ij}} \\
  & =\prod_{i             \neq            j}^{M}
  \left\{\prod_{d=1}^{D_{ij}}\prod_{n=1}^{N_{ij}^{d}}\prod_{k=1}^{K}p(W_{ij}^{dn}|
    \beta_{k})^{Z_{ij}^{dnk}}\right\}^{A_{ij}}, \\
  \end{aligned}
\end{equation*}
and 
\begin{equation*}
  \begin{aligned}
    p(Z|A, Y,\theta) &= \prod_{i \neq j}^{M}
  \left\{\prod_{d=1}^{D_{ij}}\prod_{n=1}^{N_{ij}^{d}}p(Z_{ij}^{dn}|Y_{i}, Y_{j},
    \theta)\right\}^{A_{ij}} \\
  &= \prod_{i             \neq            j}^{M}
  \left\{\prod_{d=1}^{D_{ij}}\prod_{n=1}^{N_{ij}^{d}}\prod_{q,r}^{Q}p(Z_{ij}^{dn}|\theta_{qr})^{Y_{iq}Y_{jr}}\right\}^{A_{ij}},
  \end{aligned}
\end{equation*}
as well as 
\begin{equation*}
  p(\theta)=\prod_{q,r}^{Q}\mathrm{Dir}(\theta_{qr};\alpha).
\end{equation*}

\subsection{Link with LDA and SBM}\label{ssection:probModel}

\begin{figure}
\begin{centering}
\includegraphics[width=0.5\columnwidth]{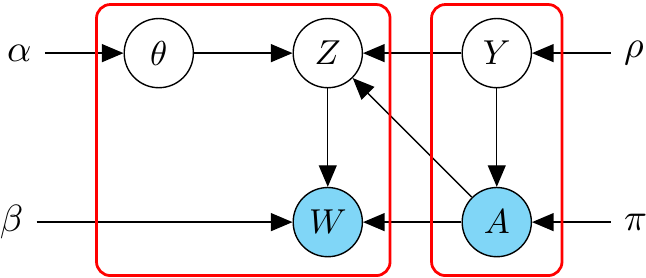}
\par\end{centering}
\protect\caption{Graphical representation of the stochastic topic block model.\label{fig:graphMod}}
\end{figure}

The full joint distribution of the STBM model is given by 
\begin{equation}\label{eq:fullcomplete}
  p(A, W, Y, Z, \theta|\rho,\pi, \beta) = p(W,Z,\theta|A, Y,\beta)p(A,Y|\rho,\pi),
\end{equation}
and  the   corresponding  graphical   model  is  provided   in  Figure
\ref{fig:graphMod}. Thus, all the documents in $W$ are involved in the
full joint distribution  through $p(W,Z,\theta|A, Y,\beta)$. Now,
let us assume that $Y$ is available. It then possible to reorganize the
documents   in   $W$   such   that   $W=(\tilde{W}_{qr})_{qr}$   where
$\tilde{W}_{qr}=\left\{W_{ij}^{d},     \forall      (d,     i,     j),
  Y_{iq}Y_{jr}A_{ij}=1\right\}$ is the set  of all documents exchanged
from  any vertex  $i$ in  cluster  $q$ to  any vertex  $j$ in  cluster
$r$. As mentioned in the previous section, each word $W_{ij}^{dn}$ has a mixture
distribution over topics which only depends on the clusters of $i$ and
$j$. Because all words in $\tilde{W}_{qr}$ are associated with the same
pair   $(q,   r)$  of   clusters,   they   share  the   same   mixture
distribution.  Removing   temporarily  the  knowledge  of   $(q,  r$),
\emph{i.e.}  simply seeing  $\tilde{W}_{qr}$  as a  document $d$,  the
sampling scheme described previously then  corresponds to the one of a
LDA model  with $D=Q^2$  independent documents  $\tilde{W}_{qr}$, each
document   having    its   own    vector   $\theta_{qr}$    of   topic
proportions. The model is then characterized by the matrix $\beta$ of
 probabilities.  Note   that  contrary  to  the   original  LDA  model
 \citep{LDA},   the  Dirichlet   distributions   considered  for   the
 $\theta_{qr}$ depend on a fixed vector $\alpha$.

As    mentioned   in    Section   \ref{ssection:presEdges},    the   second    part   of
Equation (\ref{eq:fullcomplete}) involves the sampling of the clusters and the
construction  of binary  variables  describing the  presence of  edges
between pairs of vertices. Interestingly, it corresponds exactly to the complete data
likelihood  of the  SBM model,  as considered  in \cite{zanghi08}  for
instance. Such  a likelihood term  only involves the  model parameters
$\rho$ and~$\pi$. 

\section{Inference}

We aim at maximizing the complete data log-likelihood
\begin{equation}\label{eq:completedata}
  \log p(A, W,  Y|\rho, \pi, \beta) =  \log \sum_{Z}\int_{\theta} p(A,
  W, Y, Z, \theta|\rho, \pi, \beta)d\theta,
\end{equation}
with respect to the model parameters  $(\rho, \pi, \beta)$ and the set
$Y=(Y_1,\dots,Y_M)$ of  cluster membership  vectors. Note that  $Y$ is
not  seen  here as  a  set  of latent  variables  over  which the  log-likelihood  should  be  integrated  out, as  in  standard  expectation
maximization  (EM)   \citep{dempster1977maximum}  or   variational  EM
algorithms \citep{hathaway1986another}. Moreover, the goal is not to provide any
approximate posterior  distribution of  $Y$ given  the data  and model
parameters. Conversely, $Y$ is seen here as a set of (binary) vectors for
which we aim at providing estimates. This choice is motivated by the
key property of the STBM model, \emph{i.e.}  for a given $Y$, the full
joint  distribution factorizes  into  a  LDA like  term  and SBM  like
term. In particular, given $Y$, 
words in $W$ can be seen as  being drawn from a LDA model with $D=Q^2$
documents (see Section \ref{ssection:probModel}), for which fast 
optimization  tools  have  been  derived, as pointed  out in the introduction. Note  that the choice
of optimizing a  complete data log-likelihood with respect  to the set
of cluster membership  vectors has been considered  in the literature,
for simple
  mixture model such as Gaussian mixture  models, but also for the SBM
  model \citep{zanghi08}. The
corresponding   algorithm,   so   called   classification   EM   (CEM)
\citep{celeux1991} alternates  between the  estimation of $Y$  and the
estimation of the 
model parameters. 

As mentioned previously, we introduce  our methodology in the directed
case. However, we emphasize that the  STBM package for R we developed,
implements the inference strategy for both directed and undirected networks.

\subsection{Variational decomposition}

Unfortunately, in our case, Equation (\ref{eq:completedata})
is not tractable.  Moreover the posterior distribution $p(Z, \theta|A,
W, Y, \rho, \pi, \beta)$ does not have any analytical form. Therefore, following the work of \cite{LDA} on the
LDA model, we  propose to rely on a variational  decomposition. In the
case of the STBM model, it leads to
\begin{equation*}
  \log p(A,W,Y | \rho, \pi, \beta) = \mathcal{L}\left(R(\cdot); Y, \rho,
    \pi,  \beta\right)+\mathrm{KL}\left(R(\cdot)\parallel p(\cdot|A,  W, Y,
  \rho, \pi, \beta\right)), 
\end{equation*}
where 
\begin{equation}\label{eq:lowerBound}
\mathcal{L}\left(R(\cdot); Y, \rho,
    \pi, \beta\right) =\sum_{Z}\int_{\theta}R(Z,\theta)
\log \frac{p(A,W, Y, Z, \theta|\rho, \pi, \beta)}{R(Z,\theta)} d\theta,
\end{equation}
and $\mathrm{KL}$ denotes the  Kullback-Leibler divergence between the
true and approximate posterior  distribution $R(\cdot)$ of  $(Z,\theta)$, given
the data and model parameters
\begin{equation*}
 \mathrm{KL}\left(R(\cdot)\parallel p(\cdot|A,  W, Y,
  \rho, \pi, \beta\right)) =-\sum_{Z}\int_{\theta}R(Z,\theta)\log\frac{p(Z,\theta|A,W,Y,\rho,\pi,\beta)}{R(Z,\theta)}d\theta.
\end{equation*}
Since  $ \log  p(A,W,Y | \rho, \pi, \beta)$ does  not depend on the
distribution $R(Z,\theta)$,  maximizing the lower  bound $\mathcal{L}$
with respect to $R(Z,\theta)$
induces  a minimization  of the  KL divergence.  As in  \cite{LDA}, we
assume that $R(Z, \theta)$ can be factorized over the latent variables 
in $\theta$ and $Z$. In our case, this translates into
\begin{equation*}
  R(Z, \theta) =R(Z)R(\theta)
  = R(\theta)\prod_{i \neq j, A_{ij}=1}^{M}\prod_{d=1}^{D_{ij}}\prod_{n=1}^{N_{ij}^{d}}R(Z_{ij}^{dn}).
\end{equation*}

\subsection{Model decomposition}

As pointed out in Section  \ref{ssection:probModel}, the set of latent
variables  in  $Y$   allows  the  decomposition  of   the  full  joint
distribution in two terms, from the sampling of $Y$ and $A$ to the
construction of documents  given $A$ and $Y$. When  deriving the lower
bound (\ref{eq:lowerBound}), this property leads to 
\begin{equation*}
  \mathcal{L}\left(R(\cdot); Y, \rho,
    \pi, \beta\right) = \tilde{\mathcal{L}}\left(R(\cdot); Y,
    \beta\right) + \log p(A, Y|\rho, \pi), 
\end{equation*}
where 
\begin{equation}\label{eq:tlowerBound}
  \tilde{\mathcal{L}}\left(R(\cdot);
    Y, \beta\right) = \sum_{Z}\int_{\theta}R(Z,\theta)
\log \frac{p(W, Z, \theta|A, Y,\beta)}{R(Z,\theta)} d\theta,
\end{equation}
and $ \log  p(A, Y|\rho, \pi)$ is the complete  data log-likelihood of
the  SBM  model. The  parameter  $\beta$  and the  distribution  $R(Z,
\theta)$ are only involved in the lower
bound $\tilde{\mathcal{L}}$  while $\rho$ and  $\pi$ only appear  in $
\log p(A, Y|\rho, \pi)$. Therefore, given  $Y$, these two terms can be
maximized independently. Moreover, given $Y$, $\tilde{\mathcal{L}}$ is
the lower bound  for the LDA model, as proposed  by \citet{LDA}, after
building the set $W = (\tilde{W}_{qr})_{qr}$ of $D=Q^2$ documents, as
described in Section \ref{ssection:probModel}. In the next section, we
derive a VEM algorithm  to maximize $\tilde{\mathcal{L}}$ with respect
$\beta$ and $R(Z,  \theta)$, which essentially corresponds  to the VEM
algorithm of \cite{LDA}. Then, $  \log p(A, Y|\rho, \pi)$ is maximized
with  respect  to $\rho$  and  $\pi$  to provide  estimates.  Finally,
$\mathcal{L}\left(R(\cdot); Y, \rho,
    \pi, \beta\right)$ is maximized with  respect to $Y$, which is the
  only term involved in both $\tilde{\mathcal{L}}$ and the SBM complete
  data log-likelihood.  Because the methodology we  propose requires a
  variational EM approach as well as a classification step, to provide
 estimates of $Y$, we call the corresponding strategy a classification
 VEM (C-VEM) algorithm.

\subsection{Optimization}

In  this  section, we  derive  the  optimization  steps of  the  C-VEM
algorithm  we  propose,  which  aims  at  maximizing  the  lower  bound
$\mathcal{L}$.  The algorithm alternates between the optimization of $R(Z,
 \theta)$, $Y$ and $(\rho, \pi, \beta)$ until convergence of the lower
 bound. 

\paragraph{Estimation of $R(Z,\theta)$ }

The following propositions  give the update formulae of the  E step of
the VEM algorithm applied on Equation (\ref{eq:tlowerBound}). 

\begin{proposition}
(Proof in Appendix ~\ref{app:rz})
The VEM update step for each distribution $R(Z_{ij}^{dn})$ is given by
\begin{equation*}
  R(Z_{ij}^{dn})=\mathcal{M}\left(Z_{ij}^{dn};1,\phi_{ij}^{dn}=(\phi_{ij}^{dn1},\dots, \phi_{ij}^{dnK})\right),
\end{equation*}
where 
\begin{equation*}
 \phi_{ij}^{dnk}     \propto     \left(\prod_{v=1}^{V}
  \beta_{kv}^{W_{ij}^{dnv}}\right)\prod_{q,r}^{Q}\exp\Big(\psi(\gamma_{qrk}-\psi(\sum_{l=1}^{K}\gamma_{qrl})\Big)^{Y_{iq}Y_{jr}},
\forall
(d, n, k).
\end{equation*}
$\phi_{ij}^{dnk}$ is the (approximate) posterior distribution of
words $W_{ij}^{dn}$ being in topic $k$. 
\end{proposition}

\begin{proposition}
(Proof in Appendix ~\ref{app:rtheta})
  The VEM update step for distribution $R(\theta)$ is given by
\begin{equation*}
  \begin{aligned}
  R(\theta)                                                          =
  \prod_{q,r}^{Q}\mathrm{Dir}(\theta_{qr};\gamma_{qr}=(\gamma_{qr1},\dots,
  \gamma_{qrK})),
  \end{aligned}
\end{equation*}
where 
\begin{equation*}
  \gamma_{qrk} = \alpha_{k} +   \sum_{i    \neq    j}^{M}
  A_{ij}Y_{iq}Y_{jr}\sum_{d=1}^{N_{ij}^{d}}\sum_{n=1}^{N_{ij}^{dn}}\phi_{ij}^{dnk},
  \forall (q, r, k).
\end{equation*}
\end{proposition}


\paragraph{Estimation of the model parameters}

Maximizing the lower bound $\mathcal{L}$ in Equation (\ref{eq:tlowerBound}) is used to provide estimates of the model
 parameters  $(\rho, \pi,  \beta)$.  We recall  that  $\beta$ is  only
 involved in $\tilde{\mathcal{L}}$ while  $(\rho, \pi)$ only appear in
 the SBM complete data log-likelihood. The derivation of $\tilde{\mathcal{L}}$ is given in Appendix \ref{app:bound}.
\begin{proposition} (Proofs in Appendices ~\ref{app:beta}, \ref{app:rho}, \ref{app:pi})
  The  estimates  of   $\beta$,  $\rho$,  and  $\pi$,   are  given  by
\begin{equation*}
  \beta_{kv}\propto \sum_{i \neq
  j}^{M}A_{ij}\sum_{d=1}^{D_{ij}}\sum_{n=1}^{N_{ij}^{dn}}\phi_{ij}^{dnk}W_{ij}^{dnv},
\forall (k, v),
\end{equation*}
\begin{equation*}
  \rho_{q} \propto \sum_{i=1}^{M}Y_{iq}, \forall q,
\end{equation*}
\begin{equation*}
  \pi_{qr} = \frac{   \sum_{i        \neq
    j}^{M}\sum_{q,r}^{Q}Y_{iq}Y_{jr}A_{ij}}{   \sum_{i        \neq
    j}^{M}\sum_{q,r}^{Q}Y_{iq}Y_{jr}}, \forall (q,r).
\end{equation*}
\end{proposition}

\paragraph{Estimation of $Y$}

At this step, the model parameters $(\rho, \pi, \beta)$ along with the
distribution $R(Z, \theta)$ are held fixed.  Therefore, the lower bound
$\mathcal{L}$ in  (\ref{eq:tlowerBound}) only involves the  set $Y$ of
cluster  membership  vectors. Looking  for  the  optimal solution  $Y$
maximizing this  bound is not  feasible since it involves  testing the
$Q^M$ possible cluster assignments.  However, heuristics are available
to  provide local  maxima  for this  combinatorial  problem. These  so
called \emph{greedy} methods have been used for instance to look for communities in
networks by  \cite{articlenewman2004,blondel08} but  also for  the SBM
model \citep{come15}. They are sometimes referred to as \emph{on line}
clustering methods \citep{zanghi08}. 

The algorithm  cycles randomly through  the vertices. At each  step, a
single vertex  is considered  and all  membership vectors  $Y_{j}$ are
held fixed, except $Y_{i}$. If $i$ is currently in cluster $q$, then
the method looks  for every possible label  swap, \emph{i.e.} removing
$i$ from  cluster $q$ and  assigning it to a  cluster $r \neq  q$. The
corresponding change in $\mathcal{L}$  is then computed. If no label
swap induces an increase in $\mathcal{L}$, then $Y_{i}$ remains
unchanged. Otherwise, the label swap  that yields the maximal increase
is applied,  and $Y_{i}$ is  changed accordingly. 

\subsection{Initialization strategy and model selection}

The C-VEM introduced in the  previous section allows the estimation of
$R(Z, \theta)$,  $Y$, as  well as  $(\rho, \pi,  \beta$), for  a fixed
number  $Q$ of  clusters and  a  fixed number  $K$ of  topics. As  any
EM-like algorithms, the C-VEM method depends on the initialization and
is only guaranteed to converge to a local optimum \citep{bilmes1998gentle}.
Strategies to tackle this issue include simulated annealing and the use of
multiple initializations \citep{biernacki2001strategies}. In this work, we choose
the latter option. Our C-VEM algorithm is run for several initializations of a k-means like
algorithm on a distance matrix
 between the vertices obtained as follows. 
 \begin{enumerate}
 \item  The  VEM algorithm  \citep{LDA}  for  LDA  is applied  on  the
   aggregation of  all documents exchanged  from vertex $i$  to vertex
   $j$, for each pair $(i, j)$ of vertices, in order to characterize
   a type of interaction from $i$ to $j$. Thus, a $M \times M$ matrix $X$ is first
   built such that $X_{ij}=k$ if $k$ is the majority topic used by $i$
   when discussing with $j$.
\item The  distance $M \times M$  matrix $\Delta$ is then  computed as
  follows
\begin{equation}\label{eq:dist}
\Delta(i,j)=\sum_{h=1}^{N}\delta(X_{ih}\neq X_{jh})A_{ih}A_{jh} + \sum_{h=1}^{N}\delta(X_{hi}\neq X_{hj})A_{hi}A_{hj}.
\end{equation}
The first term looks at all  possible edges from $i$ and $j$ towards a
third  vertex  $h$.  If  both  $i$  and  $j$  are  connected  to  $h$,
\emph{i.e.}  $A_{ih}A_{jh}=1$,  the  edge  types $X_{ih}$  and  $X_{jh}$  are
compared. By  symmetry, the  second term looks  at all  possible edges
from a vertex $h$ to both $i$ as well as $j$, and compare their types. Thus, the distance computes the
number of discordances in the way both $i$ and $j$ connect to other vertices or
vertices connect to them. 
 \end{enumerate}

Regarding model selection, since a model
based  approach is  proposed here,  two STBM  models will  be seen  as
different if they have different values of $Q$ and/or $K$. Therefore, the task of
estimating  $Q$   and  $K$  can   be  viewed  as  a   model  selection
problem.  Many model  selection  criteria have  been  proposed in  the
literature,    such    as    the    Akaike    information    criterion
\citep{proceedingsakaike1973}  (AIC)  and   the  Bayesian  information
criterion  \citep{Schwarz}   (BIC).   In   this  paper,   because  the
optimization  procedure considered  involves the  optimization of  the
binary matrix $Y$, we rely on a ICL-like criterion.  This criterion was originally proposed by
\cite{articlebiernacki2000} for Gaussian mixture models. In the STBM context, it
aims  at approximating  the  integrated  complete data  log-likelihood
$\log    p(A,    W,   Y)$.   

\begin{proposition}
(Proof  in   Appendix  ~\ref{app:ICL})  A
$ICL$ criterion for the STBM model can be obtained
\begin{equation*}
  ICL_{STBM} = \tilde{\mathcal{L}}(R(\cdot);Y,     \beta)    -
  \frac{K(V-1)}{2}\log Q^2 +  \max_{\rho, \pi}  \log p(A,Y|\rho,  \pi, Q)  - \frac{Q^2}{2}\log
  M(M-1) - \frac{Q-1}{2}\log M
\end{equation*}
\end{proposition}
Notice that this result relies on two  Laplace
  approximations,  a  variational  estimation,  as  well  as  Stirling
  formula. It is also worth  noticing that this criterion involves two
  parts, as shown in the appendix: a BIC like term associated to a LDA model
  \citep[see][for instance]{than2012} with $Q^2$ documents and the ICL criterion for the SBM model, as introduced by \cite{daudin2008mixture}.

\section{Numerical experiments}

This section aims at highlighting the main features of the proposed
approach on synthetic data and at proving the validity of the inference
algorithm presented in the previous section. Model selection is also
considered to validate the criterion choice. Numerical comparisons
with state-of-the-art methods conclude this section. 

\subsection{Experimental setup}

\begin{figure}
\begin{centering}
\begin{tabular}{ccc}
\includegraphics[width=0.32\columnwidth]{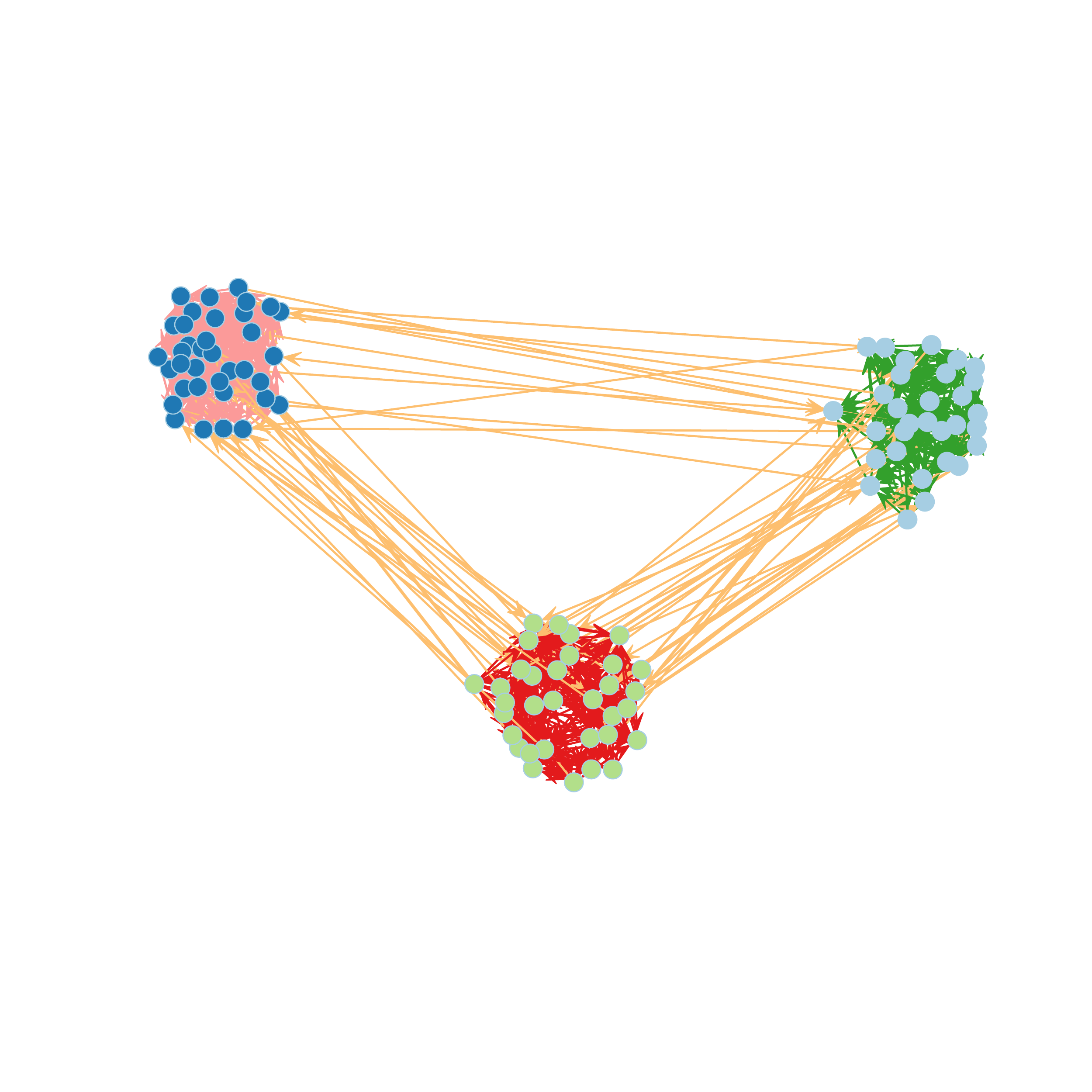} & \includegraphics[width=0.32\columnwidth]{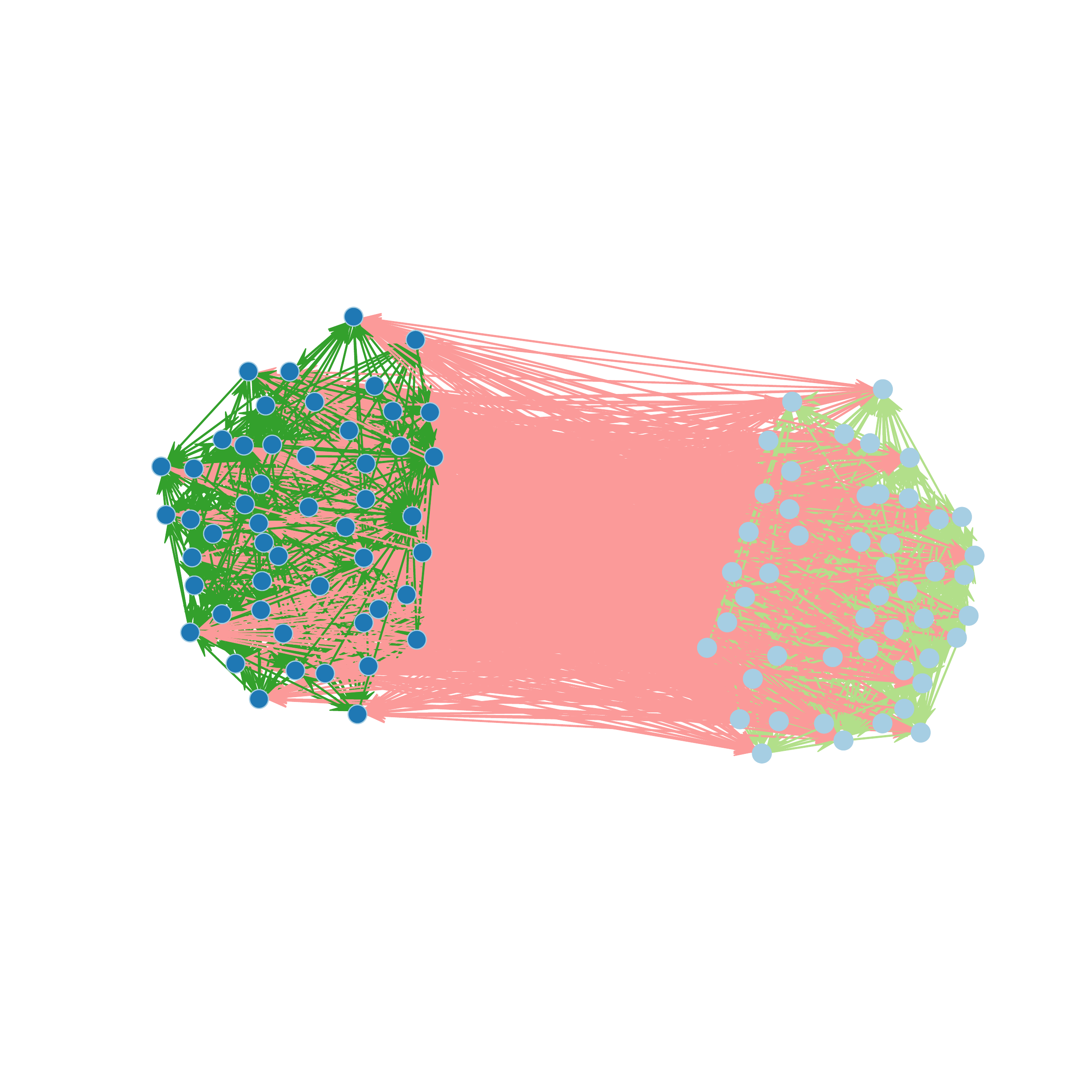} & \includegraphics[width=0.32\columnwidth]{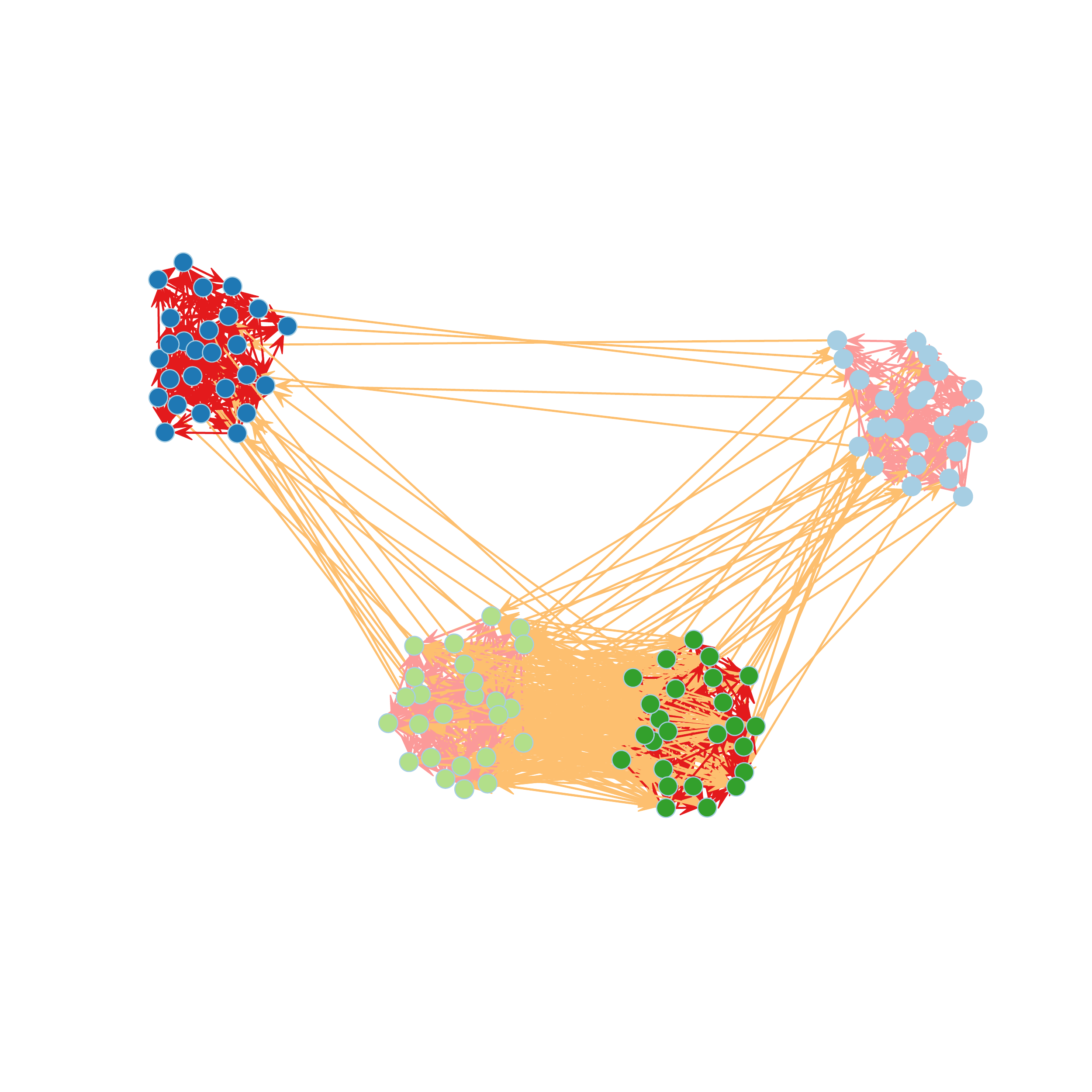}\tabularnewline
Scenario A & Scenario B & Scenario C\tabularnewline
\end{tabular}
\par\end{centering}

\protect\caption{\label{fig:Scenarios}Networks sampled according to the three simulation
scenarios A, B and C. See text for details.}
\end{figure}

\begin{table}
\begin{centering}
\begin{tabular}{|l|c|c|c|}
\hline 
Scenario & A & B & C \tabularnewline
\hline 
\hline 
M (nb of nodes) & \multicolumn{3}{c|}{100}\tabularnewline
\hline 
K (topics) & 4 & 3 & 3\tabularnewline
\hline 
Q (groups) & 3 & 2 & 4\tabularnewline
\hline 
$\rho$ (group prop.) & \multicolumn{3}{c|}{$(1/Q,...,1/Q)$}\tabularnewline
\hline 
$\pi$ (connection prob.) & $\begin{cases}
\pi_{qq}= & 0.25\\
\pi_{qr,\, r\neq q}= & 0.01
\end{cases}$ & $\pi_{qr,\,\forall q,r}=0.25$ & $\begin{cases}
\pi_{qq}= & 0.25\\
\pi_{qr,\, r\neq q}= & 0.01
\end{cases}$\tabularnewline
\hline 
$\theta$ (prop. of topics) & $\begin{cases}
\theta_{111}=\theta_{222}= & 1\\
\theta_{333}= & 1\\
\theta_{qr4,\, r\neq q}= & 1\\
\textnormal{otherwise} & 0
\end{cases}$ & $\begin{cases}
\theta_{111}=\theta_{222}= & 1\\
\theta_{qr3,\, r\neq q}= & 1\\
\textnormal{otherwise} & 0
\end{cases}$ & $\begin{cases}
\theta_{111}=\theta_{331}= & 1\\
\theta_{222}=\theta_{442}= & 1\\
\theta_{qr3,\, r\neq q}= & 1\\
\textnormal{otherwise} & 0
\end{cases}$\tabularnewline
\hline 
\end{tabular}
\par\end{centering}

\protect\caption{\label{tab:Parameter-values}Parameter values for the three simulation
scenarios (see text for details).}

\end{table}

First,  regarding  the  parametrization  of  our  approach,  we  chose
$\alpha_{k}=1,\forall k$ which induces a uniform distribution over the
topic proportions $\theta_{qr}$. 

Second, regarding the simulation setup and in order to illustrate the interest of the proposed methodology, three
different simulation setups will be used in this section. To simplify
the characterization and facilitate the reproducibility of the experiments,
we designed three different scenarios. They are as follows:
\begin{itemize}
\item scenario A consists in networks with $Q=3$ groups, corresponding
to clear communities, where persons within a group talk preferentially
about a unique topic and use a different topic when talking with persons
of other groups. Thus, those networks contain $K=4$ topics.
\item scenario B consists in networks with a unique community where the
$Q=2$ groups are only differentiated by the way they discuss within
and between groups. Persons within groups 1 and 2 talk preferentially
about topics 1 and 2 respectively. A third topic is used for the
communications between persons of different groups.
\item scenario C, finally, consists in networks with $Q=4$ groups which
use $K=3$ topics to communicate. Among the $4$ groups, two groups
correspond to clear communities where persons talk preferentially
about a unique topic within the communities. The two other groups
correspond to a single community and are only discriminated by the
topic used in the communications. People from group 3 use topic
1 and the topic 2 is used in group 4. The third topic is used
for communications between groups.
\end{itemize}
For all scenarios, the simulated messages are sampled from four texts
from BBC news: one text is about the birth of Princess Charlotte,
the second one is about black holes in astrophysics, the third one
is focused on UK politics and the last one is about cancer diseases
in medicine. All messages are made of 150 words. Table~\ref{tab:Parameter-values}
provides the parameter values for the three simulation scenarios.
Figure~\ref{fig:Scenarios} shows simulated networks according to
the three simulation scenarios. It is worth noticing that all simulation
scenarios have been designed such that they do not to strictly follow
the STBM model and therefore they do not favor the model we propose in comparisons.

\subsection{Introductory example}

\begin{figure}[p]
\begin{centering}
\includegraphics[width=0.5\textwidth]{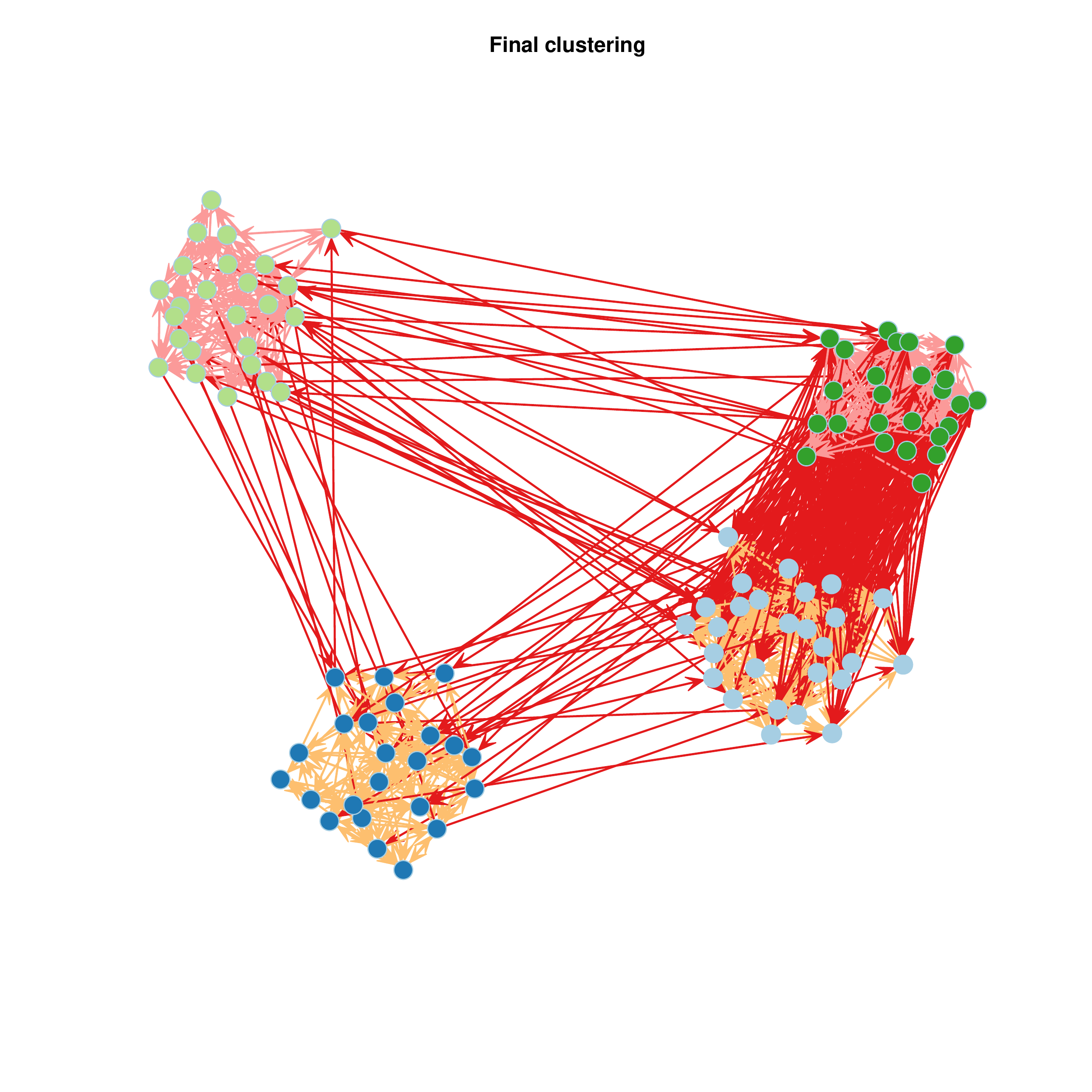}
\par\end{centering}

\protect\caption{\label{fig:Clustering-IntroEx}Clustering result for the introductory
example (scenario C). See text for details.}
\end{figure}

\begin{figure}[p]
\centering{}%
\begin{tabular}{cc}
\includegraphics[width=0.48\textwidth]{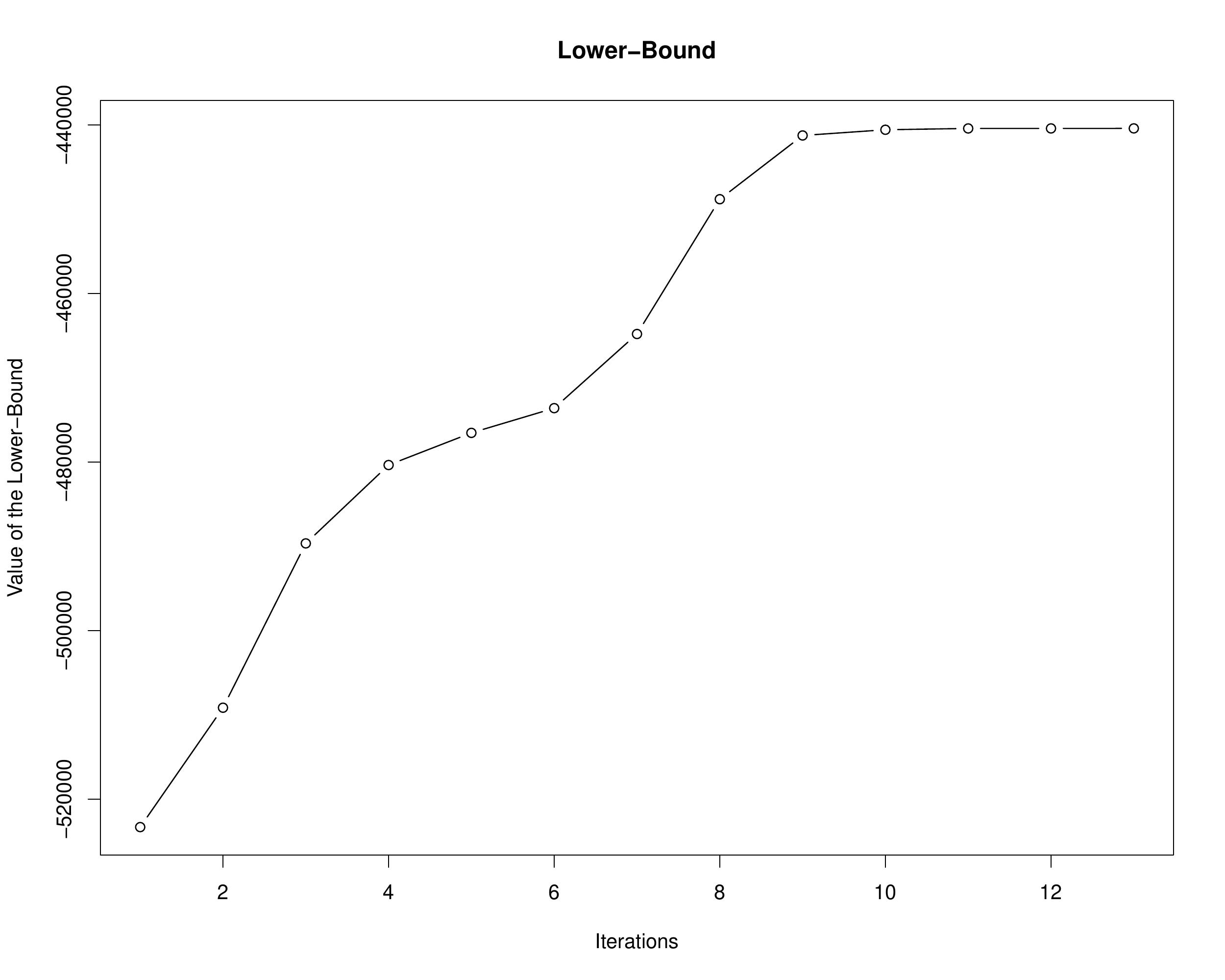} & \includegraphics[width=0.48\textwidth]{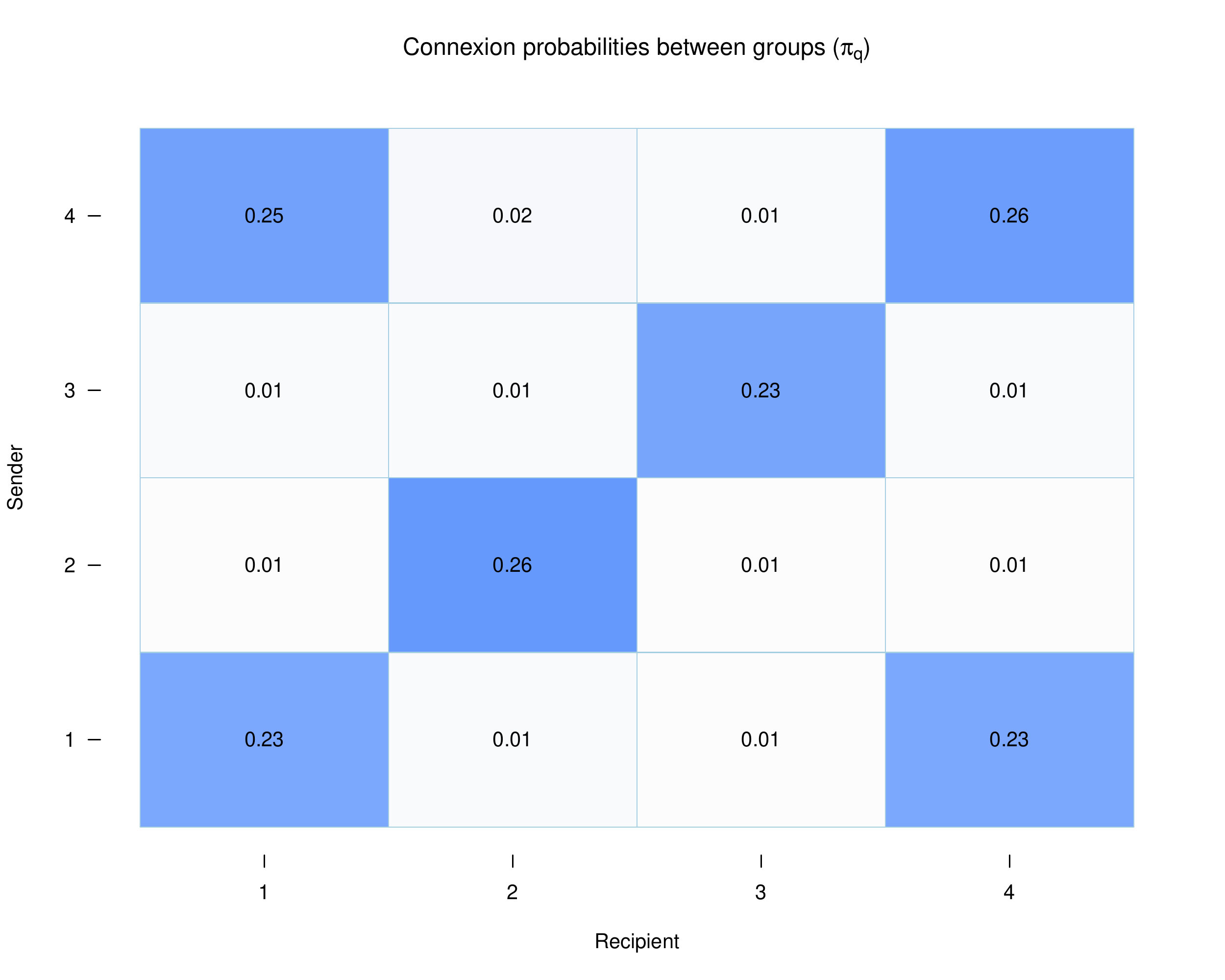}\tabularnewline
\includegraphics[width=0.48\textwidth]{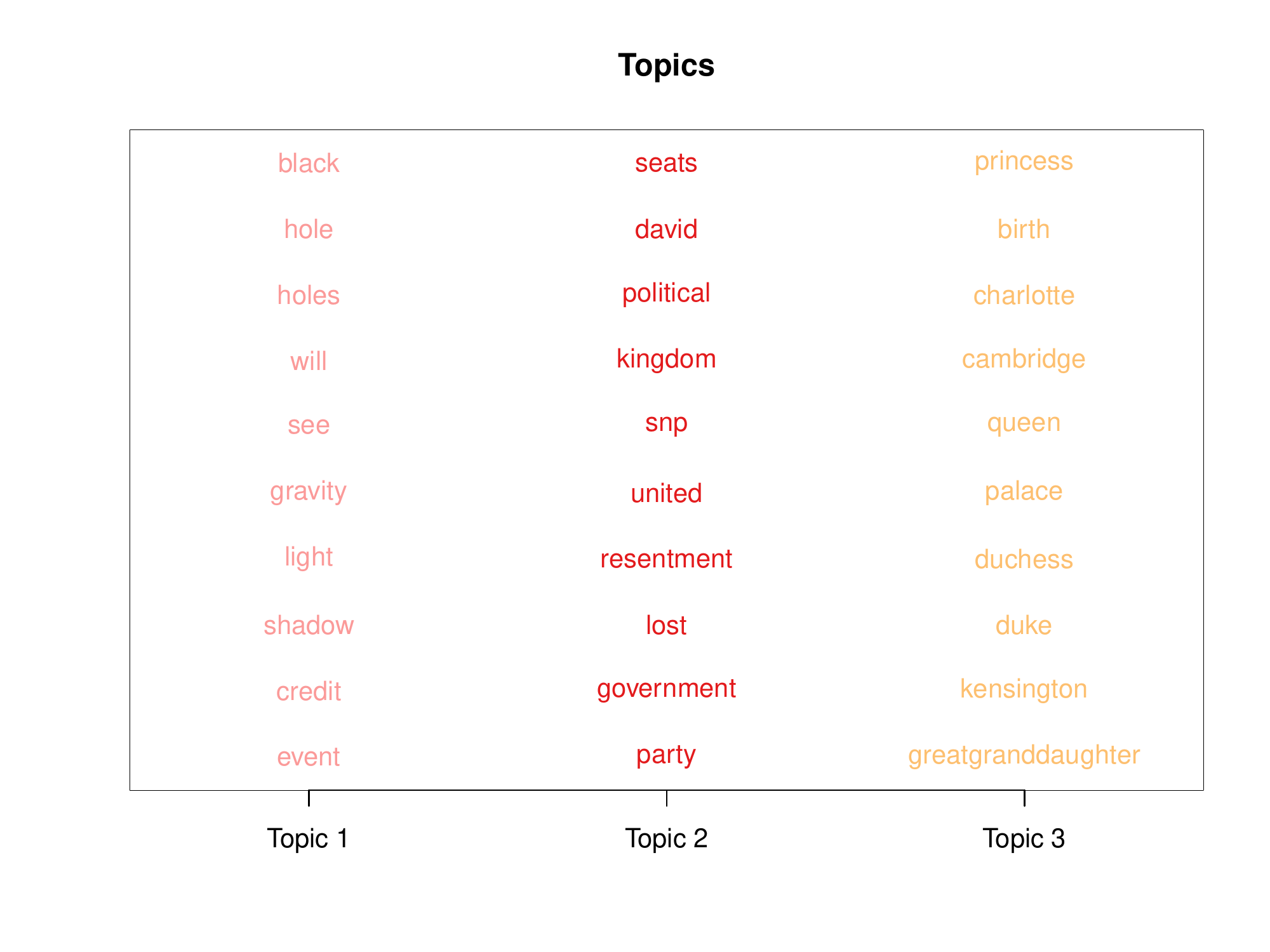} & \includegraphics[width=0.48\textwidth]{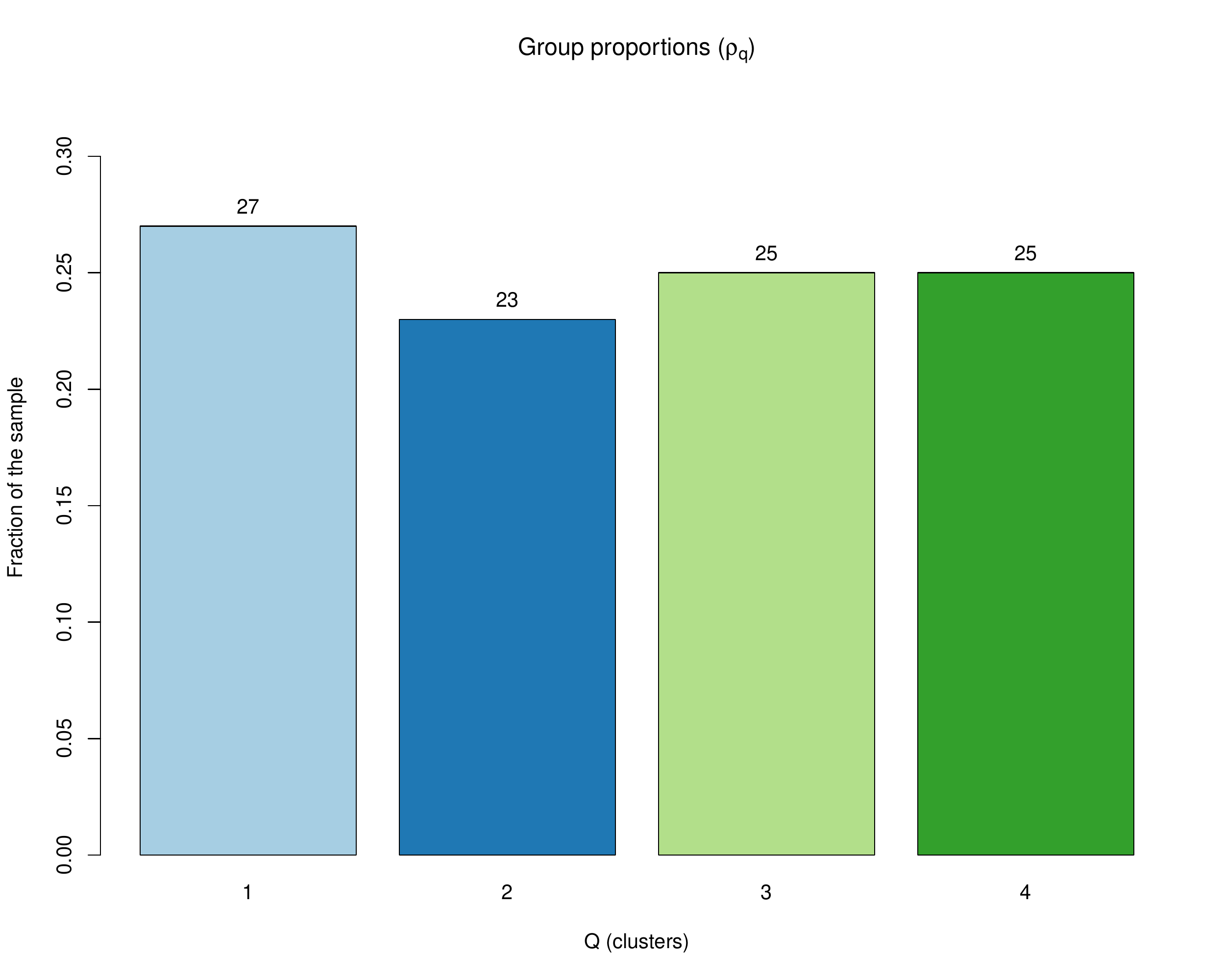}\tabularnewline
\end{tabular}\protect\caption{\label{fig:Clustering-IntroEx-1}Clustering result for the introductory
example (scenario C). See text for details.}
\end{figure}

As an introductory example, we consider a network of $M=100$ nodes
sampled according to scenario C ($3$ communities, $Q=4$ groups and
$K=3$ topics). This scenario corresponds to a situation where both
network structure and topic information are needed to correctly recover
the data structure. Indeed, groups 3 and 4 form a single community
when looking at the network structure and it is necessary to look at
the way they communicate to discriminate the two groups.

The C-VEM algorithm for STBM was run on the network with the actual
number of groups and topics (the problem of model selection will be
considered in next section). Figure~\ref{fig:Clustering-IntroEx}
first shows the obtained clustering, which is here perfect both regarding
the simulated node and edges partitions. More interestingly, Figure~\ref{fig:Clustering-IntroEx-1}
allows to visualize the evolution of the lower bound $\mathcal{L}$
along the algorithm iterations (top-left panel), the estimated model
parameters $\pi$ and $\rho$ (right panels), and the most frequent
words in the $3$ found topics (left-bottom panel). It turns out that
both the model parameters, $\pi$ and $\rho$ (see Table~\ref{tab:Parameter-values}
for actual values), and the topic meanings are well recovered. STBM
indeed perfectly recovers the three themes that we used for simulating
the textual edges: one is a ``royal baby'' topic, one is a political
one and the last one is focused on Physics. Notice also that this result
was obtained in only a few iterations of the C-VEM algorithm, that
we proposed for inferring STBM models.

\begin{figure}[t]
\begin{centering}
\includegraphics[width=0.5\columnwidth]{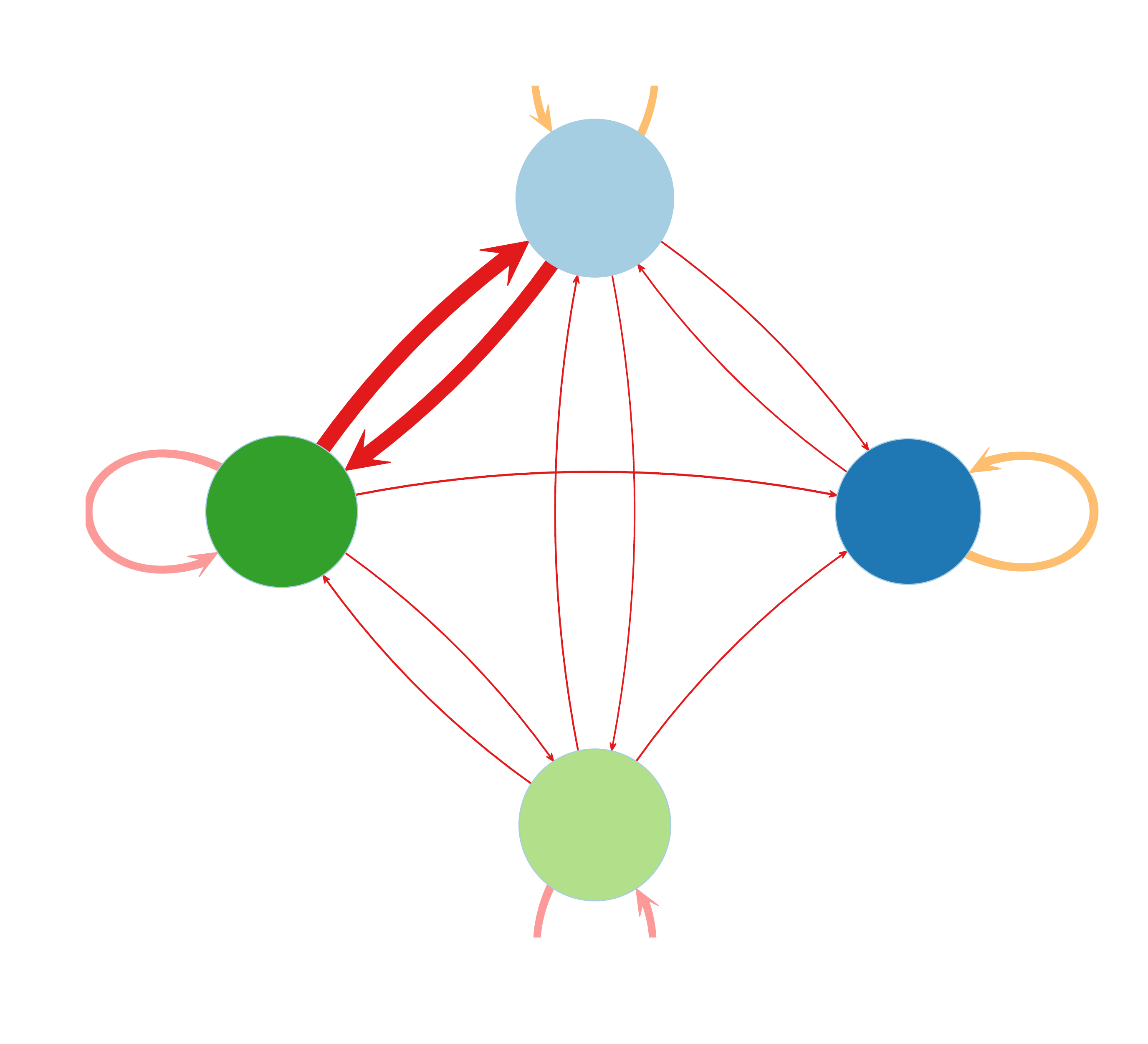}
\par\end{centering}

\protect\caption{\label{fig:Clustering-IntroEx-2}Introductory example: summary of
connexion probabilities between groups ($\pi$, edge widths), group
proportions ($\rho$, node sizes) and most probable topics for group
interactions (edge colors).}
\end{figure}

A useful and compact view of both parameters $\pi$ and $\rho$, and
of the most probable topics for group interactions can be offered
by Figure~\ref{fig:Clustering-IntroEx-2}. Here, edge widths correspond
to connexion probabilities between groups ($\pi$), the node sizes
are proportional to group proportions ($\rho$) and edge colors indicate
the majority topics for group interactions. It is important to notice
that, even though only the most probable topic is displayed here, each
textual edge may use different topics.

\subsection{Model selection}

\begin{table}
\begin{centering}
\begin{tabular}{|c|c|c|c|c|c|c|}
\hline 
\multicolumn{7}{|c|}{Scenario A ($Q=3$, $K=4$)}\tabularnewline
\hline 
\hline 
$K$\textbackslash{}$Q$ & 1 & 2 & 3\cellcolor[gray]{0.9} & 4 & 5 & 6\tabularnewline
\hline 
1 & 0 & 0 & 0\cellcolor[gray]{0.9} & 0 & 0 & 0\tabularnewline
\hline 
2 & 12 & 0 & 0\cellcolor[gray]{0.9} & 0 & 0 & 0\tabularnewline
\hline 
3 & 0 & 0 & 0\cellcolor[gray]{0.9} & 0 & 0 & 0\tabularnewline
\hline 
\rowcolor[gray]{0.9}4 & 0 & 0 & \textbf{82} & 2 & 0 & 2\tabularnewline
\hline 
5 & 0 & 0 & 2\cellcolor[gray]{0.9} & 0 & 0 & 0\tabularnewline
\hline 
6 & 0 & 0 & 0\cellcolor[gray]{0.9} & 0 & 0 & 0\tabularnewline
\hline 
\end{tabular}~~~%
\begin{tabular}{|c|c|c|c|c|c|c|}
\hline 
\multicolumn{7}{|c|}{Scenario B ($Q=2$, $K=3$)}\tabularnewline
\hline 
\hline 
$K$\textbackslash{}$Q$ & 1 & 2\cellcolor[gray]{0.9} & 3 & 4 & 5 & 6\tabularnewline
\hline 
1 & 0 & 0\cellcolor[gray]{0.9} & 0 & 0 & 0 & 0\tabularnewline
\hline 
2 & 12 & 0\cellcolor[gray]{0.9} & 0 & 0 & 0 & 0\tabularnewline
\hline 
\rowcolor[gray]{0.9}3 & 0 & \textbf{88} & 0 & 0 & 0 & 0\tabularnewline
\hline 
4 & 0 & 0\cellcolor[gray]{0.9} & 0 & 0 & 0 & 0\tabularnewline
\hline 
5 & 0 & 0\cellcolor[gray]{0.9} & 0 & 0 & 0 & 0\tabularnewline
\hline 
6 & 0 & 0\cellcolor[gray]{0.9} & 0 & 0 & 0 & 0\tabularnewline
\hline 
\end{tabular}~~~%
\begin{tabular}{|c|c|c|c|c|c|c|}
\hline 
\multicolumn{7}{|c|}{Scenario C ($Q=4$, $K=3$)}\tabularnewline
\hline 
\hline 
$K$\textbackslash{}$Q$ & 1 & 2 & 3 & 4\cellcolor[gray]{0.9} & 5 & 6\tabularnewline
\hline 
1 & 0 & 0 & 0 & 0\cellcolor[gray]{0.9} & 0 & 0\tabularnewline
\hline 
2 & 0 & 0 & 0 & 0\cellcolor[gray]{0.9} & 0 & 0\tabularnewline
\hline 
\rowcolor[gray]{0.9}3 & 0 & 0 & 2 & \textbf{82} & 0 & 0\tabularnewline
\hline 
4 & 0 & 0 & 0 & 16\cellcolor[gray]{0.9} & 0 & 0\tabularnewline
\hline 
5 & 0 & 0 & 0 & 0\cellcolor[gray]{0.9} & 0 & 0\tabularnewline
\hline 
6 & 0 & 0 & 0 & 0\cellcolor[gray]{0.9} & 0 & 0\tabularnewline
\hline 
\end{tabular}
\par\end{centering}

\protect\caption{\label{tab:Number-of-selections}Percentage of selections by ICL for
each STBM model $(Q,K)$ on 50 simulated networks of each of three
scenarios. Highlighted rows and columns correspond to the actual values
for $Q$ and $K$.}

\end{table}

This experiment focuses on the ability of the ICL criterion to select
 appropriate values for $Q$ and $K$. To this end, we simulated
50 networks according to each of the three scenarios and STBM was
applied on those networks for values of $Q$ and $K$ ranging from
$1$ to $6$. Table~\ref{tab:Number-of-selections} presents the
percentage of selections by ICL for each STBM model $(Q,K)$ on 50
simulated networks of each of the three scenarios. 

In the  three different situations, ICL  succeeds most of the  time in
identifying the actual combination of the number of groups and topics.
For scenarios A and B, when ICL does not select the correct values
for $Q$ and $K$, the criterion seems to underestimate the values
of $Q$ and $K$ whereas it tends to overestimate them in case of
scenario C. One can also notice that wrongly selected models are usually
close to the simulated one. Let us also recall that, since the data
are not strictly simulated according to a STBM model, the ICL criterion
does not have the model which generated the data in the set of tested models. This experiment
allows to validate ICL as a model selection tool for STBM.

\subsection{Benchmark study}

This third experiment aims at comparing the ability of STBM to recover
the network structure both in term of node partition and topics. STBM
is here compared to SBM, using the mixer package~\citep{mixer},
and LDA, using the topicmodels package \citep{topicmodels}. Obviously, SBM and LDA
will be only able to recover either the node partition or the topics.
We chose here to evaluate the results by comparing the resulting node
and topic partitions with the actual ones (the simulated partitions).
In the clustering community, the adjusted Rand index (ARI)~\citep{rand1971objective}
serves as a widely accepted criterion for the difficult task of clustering
evaluation. The ARI looks at all pairs of nodes and checks whether
they are classified in the same group or not in both partitions. As
a  result, an  ARI value  close  to 1  means that  the partitions  are
similar. Notice that the actual values  of $Q$ and $K$ are provided to
the three algorithms.

In addition to the different simulation scenarios, we considered three
different situations: the standard simulation situation as described
in Table~\ref{tab:Parameter-values} (hereafter ``Easy''), a simulation
situation (hereafter ``Hard 1'') where the communities are less
differentiated ($\pi_{qq}=0.25$ and $\pi_{q\neq r}=0.2$, except
for scenario B) and a situation (hereafter ``Hard 2'') where $40\%$
of message words are sampled in different topics than the actual topic. 

\begin{table}
\begin{centering}
\begin{tabular}{|c|c|cccccc|}
\hline 
\multirow{5}{*}{\begin{turn}{90}
Easy
\end{turn}} &  & \multicolumn{2}{c}{Scenario A} & \multicolumn{2}{c}{Scenario B} & \multicolumn{2}{c|}{Scenario C}\tabularnewline
 & Method & node ARI & edge ARI & node ARI & edge ARI & node ARI & edge ARI\tabularnewline
\cline{2-8} 
 & SBM & 1.00$\pm$0.00 & -- & 0.01$\pm$0.01 & -- & 0.69$\pm$0.07 & --\tabularnewline
 & LDA & -- & 0.97$\pm$0.06 & -- & 1.00$\pm$0.00 & -- & 1.00$\pm$0.00\tabularnewline
 & STBM & 0.98$\pm$0.04 & 0.98$\pm$0.04 & 1.00$\pm$0.00 & 1.00$\pm$0.00 & 1.00$\pm$0.00 & 1.00$\pm$0.00\tabularnewline
\hline 
\end{tabular}\medskip
\par\end{centering}

\begin{centering}
\begin{tabular}{|c|c|cccccc|}
\hline 
\multirow{5}{*}{\begin{turn}{90}
Hard 1
\end{turn}} &  & \multicolumn{2}{c}{Scenario A} & \multicolumn{2}{c}{Scenario B} & \multicolumn{2}{c|}{Scenario C}\tabularnewline
 & Method & node ARI & edge ARI & node ARI & edge ARI & node ARI & edge ARI\tabularnewline
\cline{2-8} 
 & SBM & 0.01$\pm$0.01 & -- & 0.01$\pm$0.01 & -- & 0.01$\pm$0.01 & --\tabularnewline
 & LDA & -- & 0.90$\pm$0.17 & -- & 1.00$\pm$0.00 & -- & 0.99$\pm$0.01\tabularnewline
 & STBM & 1.00$\pm$0.00 & 0.90$\pm$0.13 & 1.00$\pm$0.00 & 1.00$\pm$0.00 & 1.00$\pm$0.00 & 0.98$\pm$0.03\tabularnewline
\hline 
\end{tabular}\medskip
\par\end{centering}

\begin{centering}
\begin{tabular}{|c|c|cccccc|}
\hline 
\multirow{5}{*}{\begin{turn}{90}
Hard 2
\end{turn}} &  & \multicolumn{2}{c}{Scenario A} & \multicolumn{2}{c}{Scenario B} & \multicolumn{2}{c|}{Scenario C}\tabularnewline
 & Method & node ARI & edge ARI & node ARI & edge ARI & node ARI & edge ARI\tabularnewline
\cline{2-8} 
 & SBM & 1.00$\pm$0.00 & -- & -0.01$\pm$0.01 & -- & 0.65$\pm$0.05 & --\tabularnewline
 & LDA & -- & 0.21$\pm$0.13 & -- & 0.08$\pm$0.06 & -- & 0.09$\pm$0.05\tabularnewline
 & STBM & 0.99$\pm$0.02 & 0.99$\pm$0.01 & 0.59$\pm$0.35 & 0.54$\pm$0.40 & 0.68$\pm$0.07 & 0.62$\pm$0.14\tabularnewline
\hline 
\end{tabular}
\par\end{centering}

\protect\caption{Clustering results for the SBM, LDA and STBM on 20 networks simulated
according to the three scenarios. Average ARI values are reported
with standard deviations for both node and edge clustering. The ``Easy''
situation corresponds to the simulation situation describes in Table~\ref{tab:Parameter-values}.
In the ``Hard 1'' situation, the communities are very few differentiated
($\pi_{qq}=0.25$ and $\pi_{q\protect\neq r}=0.2$, except for scenario
B). The ``Hard 2'' situation finally corresponds to a setup where
$40\%$ of message words are sampled in different topics than the
actual topic. }
\end{table}

In the ``Easy'' situation, the results are coherent with our initial
guess when building the simulation scenarios. Indeed, besides the
fact that SBM and LDA are only able to recover one of the two partitions,
scenario A is an easy situation for all methods since the clusters
perfectly match the topic partition. Scenario B, which has no communities
and where groups only depend on topics, is obviously a difficult situation
for SBM but does not disturb LDA which perfectly recovers the topics.
In scenario C, LDA still succeeds in identifying the topics whereas
SBM well recognizes the two communities but fails in discriminating
the two groups hidden in a single community. Here, STBM obtains in
all scenarios the best performance on both nodes and edges.

The ``Hard 1'' situation considers the case where the communities
are actually not well differentiated. Here, LDA is little affected (only
in scenario A) whereas SBM is no longer able to distinguish the groups
of nodes. Conversely, STBM relies on the found topics to correctly
identifies the node groups and obtains, here again, excellent ARI
values in all the three scenarios.

The last situation, the so-called ``Hard 2'' case, aims at highlighting
the effect of the word sampling in the recovering of the used topics.
On the one hand, SBM now achieves a satisfying classification of nodes
for scenarios A and C while LDA fails in recovering the majority topic
used for simulation. On those two scenarios, STBM performs well on
both nodes and topics. This proves that STBM is also able to recover
the topics in a noisy situation by relying on the network structure.
On the other hand, scenario B presents an extremely difficult situation
where topics are noised and there are no communities. Here, although
both LDA and SBM fail, STBM achieves a satisfying result on both nodes
and edges. This is, once again, an illustration of the fact that the
joint modeling of network structure and topics allows to recover complex
hidden structures in a network with textual edges.

\section{Application to real-world problems}

In this section, we present two applications of STBM to real-world
networks: the Enron email and the Nips co-authorship networks. These
two data sets have been chosen because one is a directed network of
moderate size whereas the other one is undirected and of a large size.

\subsection{Analysis of the Enron email network}

\begin{figure}[t]
\begin{centering}
\includegraphics[width=0.8\columnwidth]{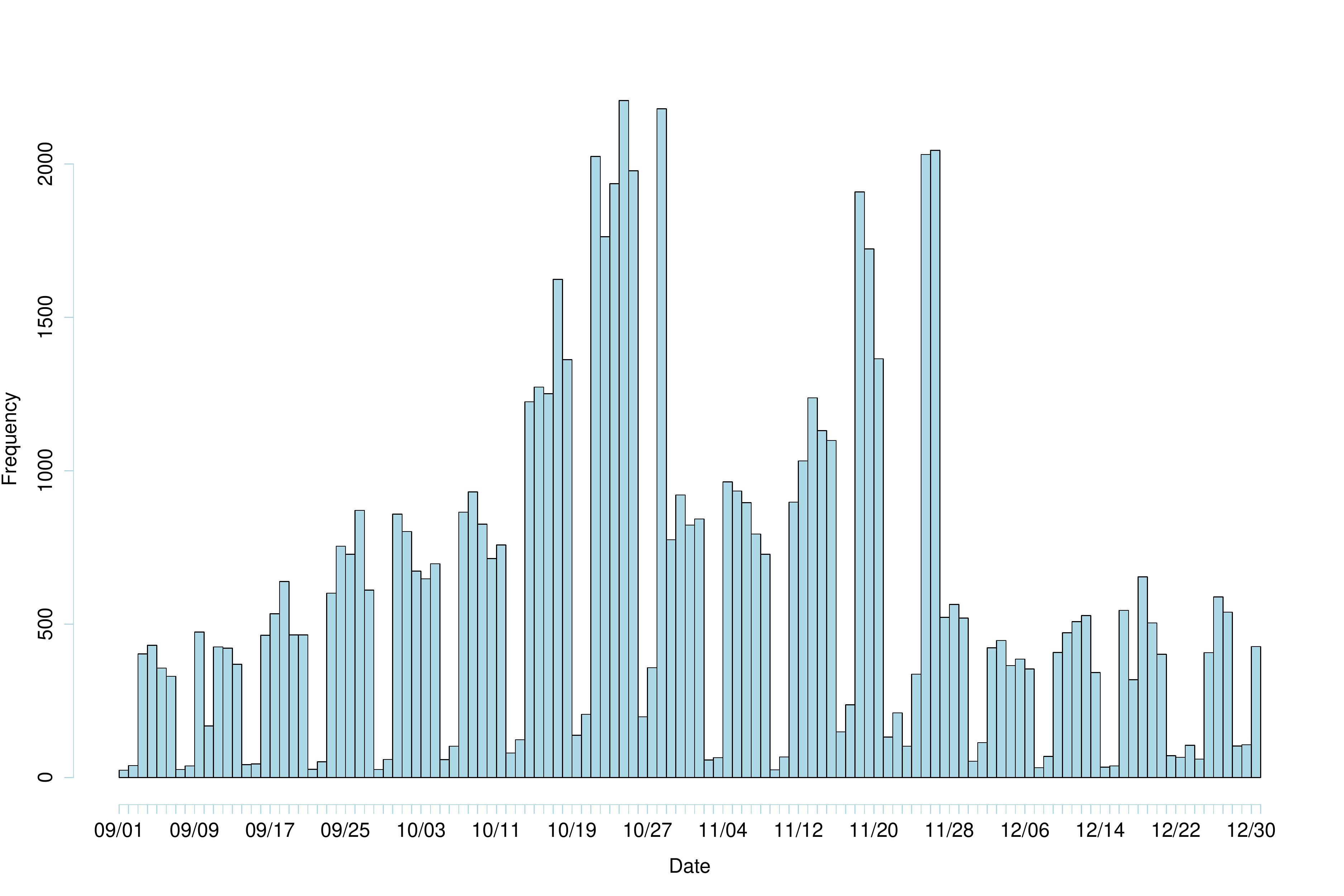}
\par\end{centering}

\protect\caption{\label{fig:Frequency-of-messages}Frequency of messages between Enron
employees between September 1st and December 31th, 2001.}
\end{figure}

We consider here a classical communication network, the Enron data
set, which contains all email communications between 149 employees
of the famous company from 1999 to 2002. The original data set is
available at \url{https://www.cs.cmu.edu/~./enron/}. Here, we focus
on the period September, 1st to December, 31th, 2001. We chose this
specific time window because it is the denser period in term of sent
emails and since it corresponds to a critical period for the company.
Indeed, after the announcement early September 2001 that the company
was ``in the strongest and best shape that it has ever been in'',
the Securities and Exchange Commission (SEC) opened an investigation
on October, 31th for fraud and the company finally filed for bankruptcy
on December, 2nd, 2001. By this time, it was the largest bankruptcy
in U.S. history and resulted in more than 4,000 lost jobs. Unsurprisingly,
those key dates actually correspond to breaks in the email activity
of the company, as shown by Figure~\ref{fig:Frequency-of-messages}.

\begin{figure}[p]
\begin{centering}
\includegraphics[width=0.75\columnwidth]{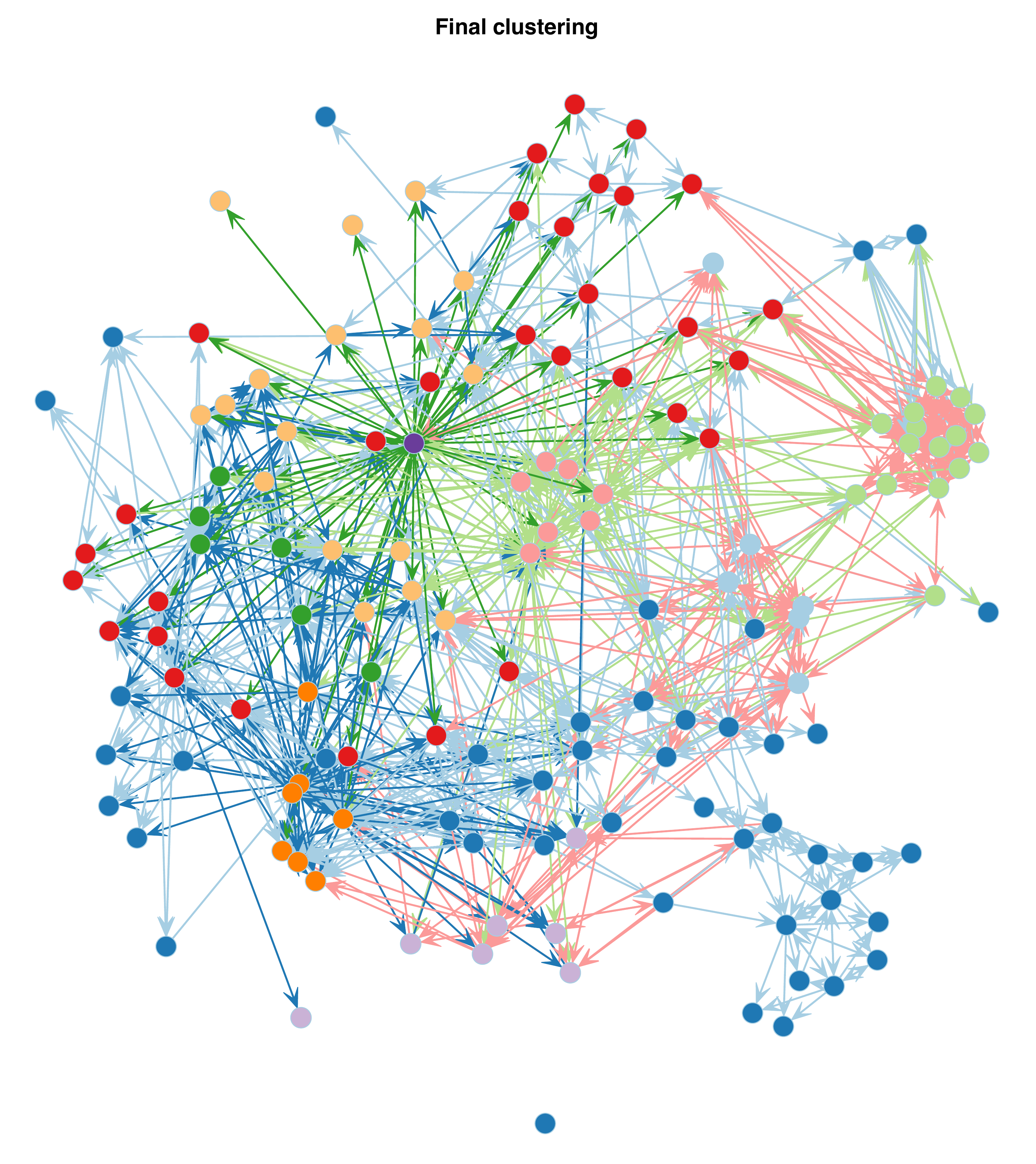}\includegraphics[width=0.25\columnwidth]{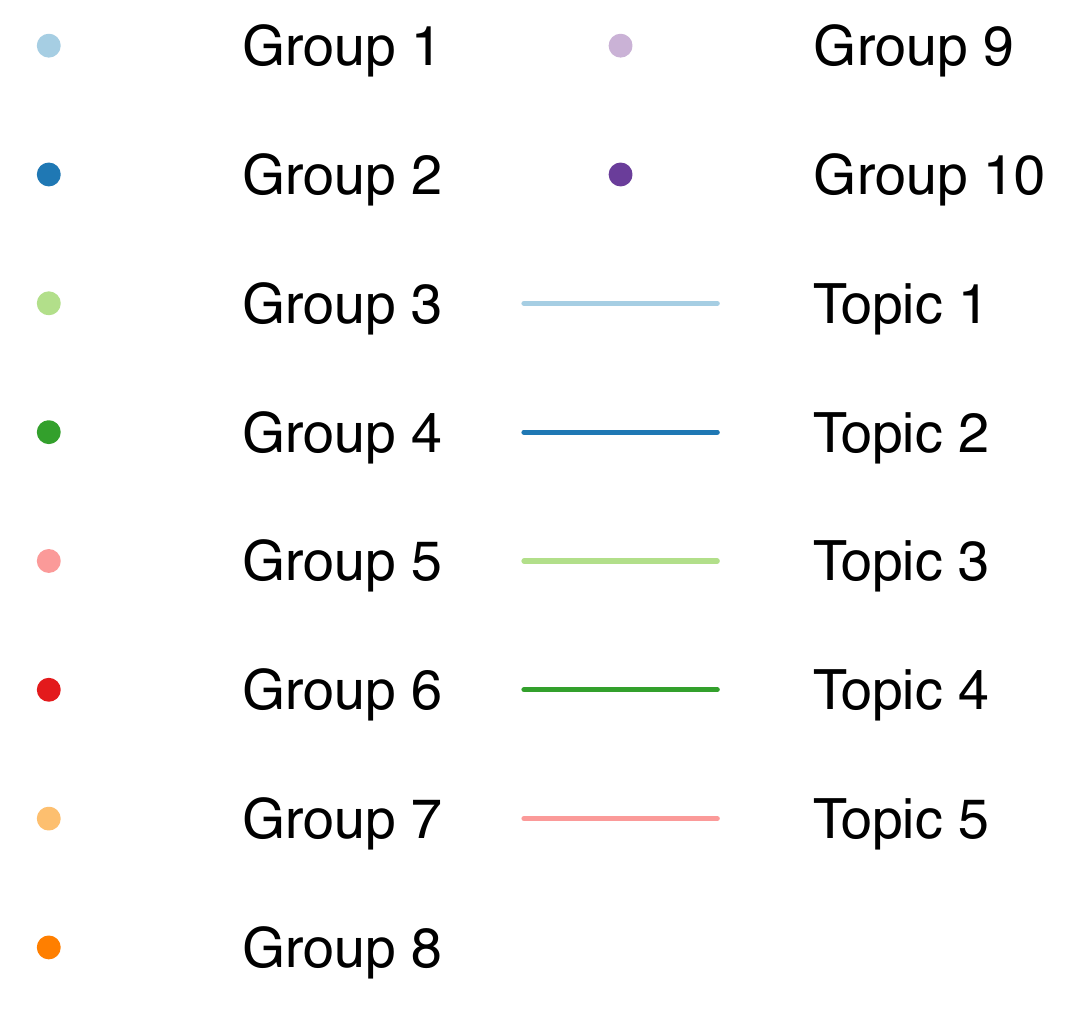}
\par\end{centering}

\protect\caption{\label{fig:Enron-Clustering-result}Clustering result with STBM on
the Enron data set (Sept.-Dec. 2001).}
\end{figure}

\begin{figure}[p]
\begin{centering}
\includegraphics[width=0.85\columnwidth]{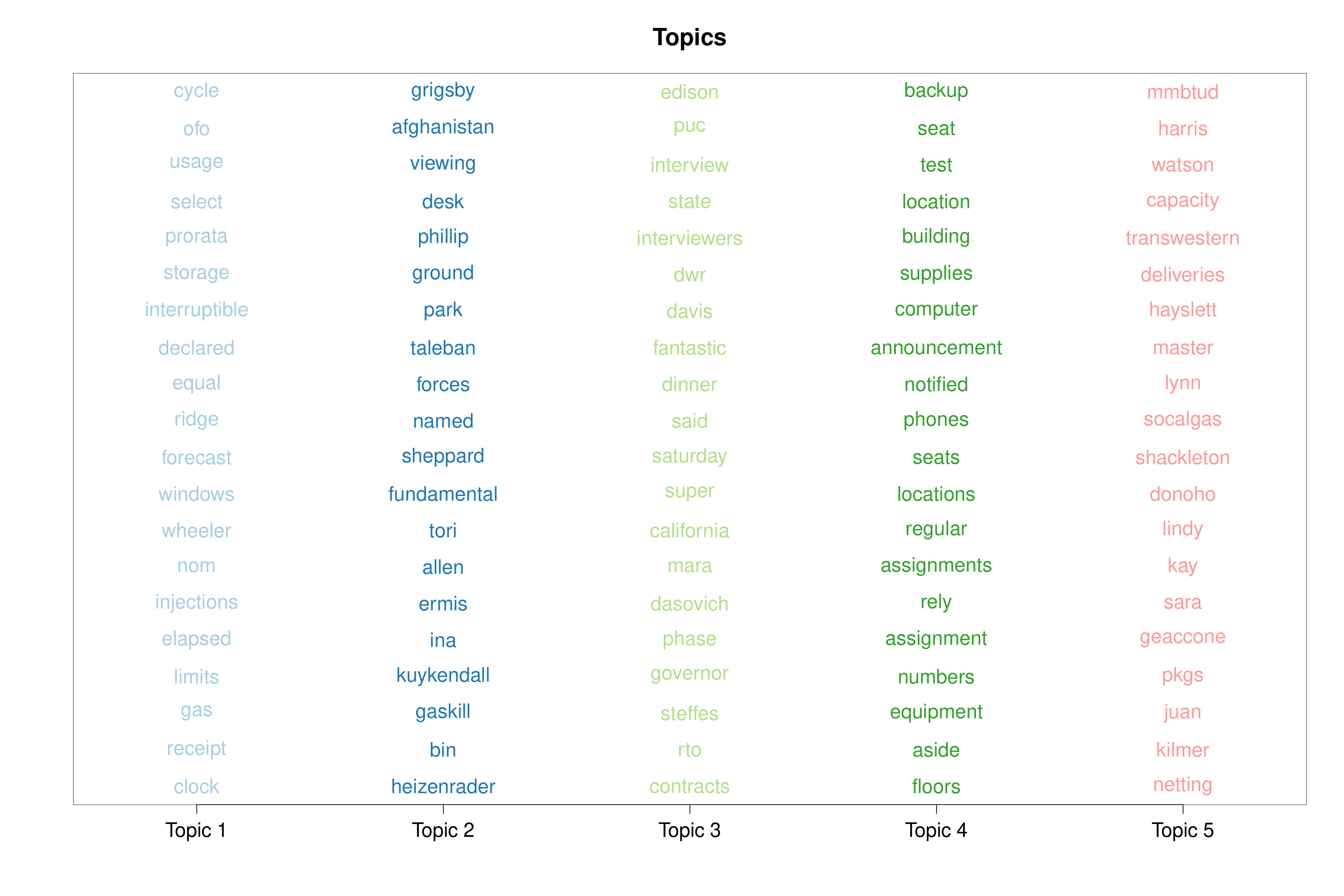}
\par\end{centering}

\protect\caption{\label{fig:Enron-Terms}Most specific words for the 5 found topics
with STBM on the Enron data set.}
\end{figure}

The data set considered here contains 20~940 emails sent between
the $M=149$ employees. All messages sent between two individuals
were coerced in a single meta-message. Thus, we end up with a data
set of 1~234 directed edges between employees, each edge carrying
the text of all messages between two persons.

The C-VEM algorithm we developed for STBM was run on these data for
a number $Q$ of groups from $1$ to $14$ and a number $K$ of topics
from $2$ to $20$. As one can see on Figure~1 of the supplementary material
the model with the highest value was $(Q,K)=(10,5)$. Figure~\ref{fig:Enron-Clustering-result}
shows the clustering obtained with STBM for $10$ groups of nodes
and $5$ topics. As previously, edge colors refer to the majority
topics for the communications between the individuals. The found topics
can be easily interpreted by looking at the most specific words of
each topic, displayed in Figure~\ref{fig:Enron-Terms}. In a few
words, we can summarize the found topics as follows:
\begin{itemize}
\item Topic 1 seems to refer to the financial and trading activities of
Enron, 
\item Topic 2 is concerned with Enron activities in Afghanistan (Enron and
the Bush administration were suspected to work secretly with Talibans
up to a few weeks before the 9/11 attacks),
\item Topic 3 contains elements related to the California electricity crisis,
in which Enron was involved, and which almost caused the bankruptcy
of SCE-corp (Southern California Edison Corporation) early 2001,
\item Topic 4 is about usual logistic issues (building equipment, computers,
...),
\item Topic 5 refers to technical discussions on gas deliveries (mmBTU represents
1 million of British thermal unit, which is equal to 1055 joules). 
\end{itemize}

\begin{figure}[t]
\begin{centering}
\includegraphics[width=0.5\columnwidth]{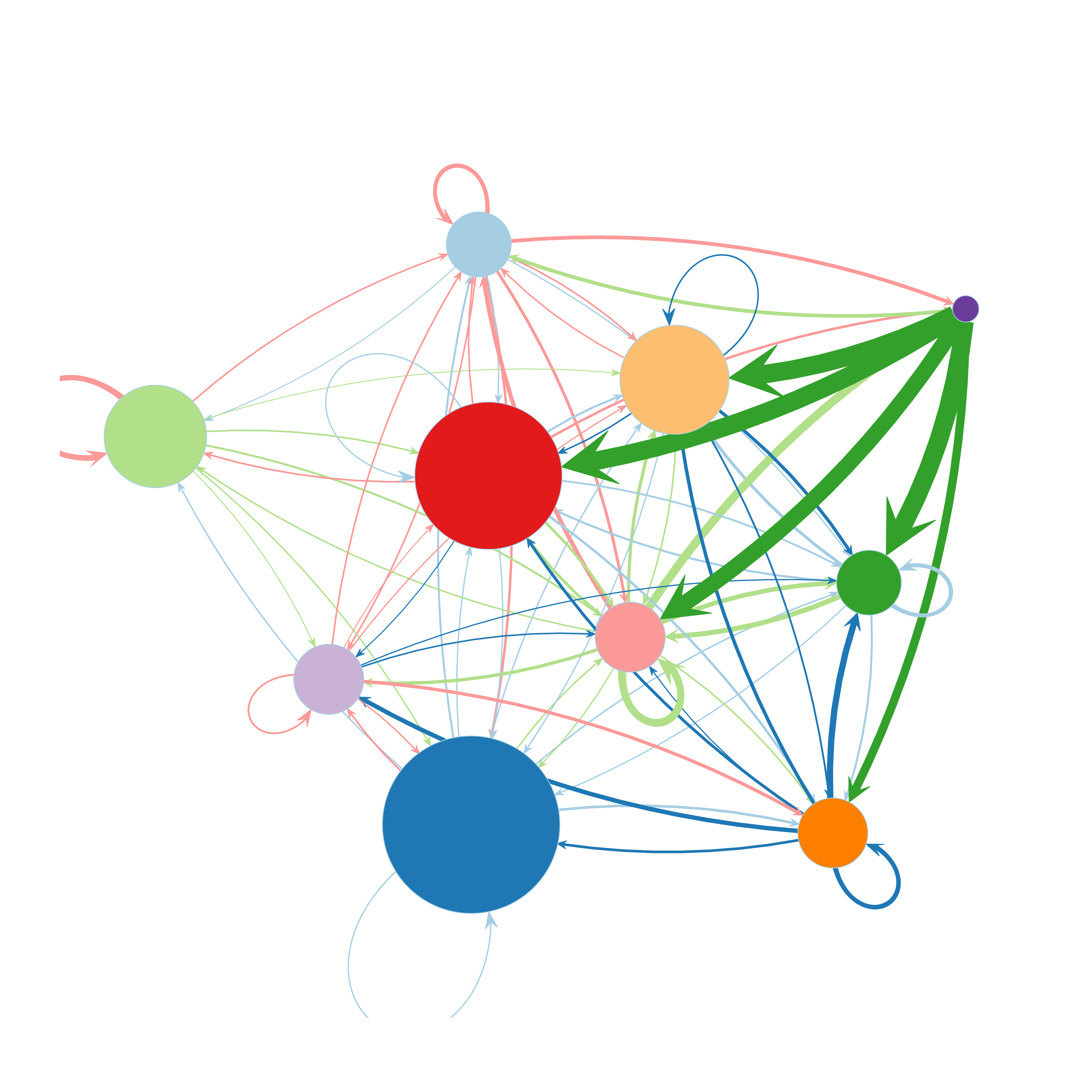}\includegraphics[width=0.25\columnwidth]{images/Enron-Network-legend.pdf}
\par\end{centering}

\protect\caption{\label{fig:Enron-metaReso}Enron data set: summary of connexion probabilities
between groups ($\pi$, edge widths), group proportions ($\rho$,
node sizes) and most probable topics for group interactions (edge
colors).}
\end{figure}

\begin{figure}[t]
\begin{centering}
\includegraphics[width=0.9\columnwidth]{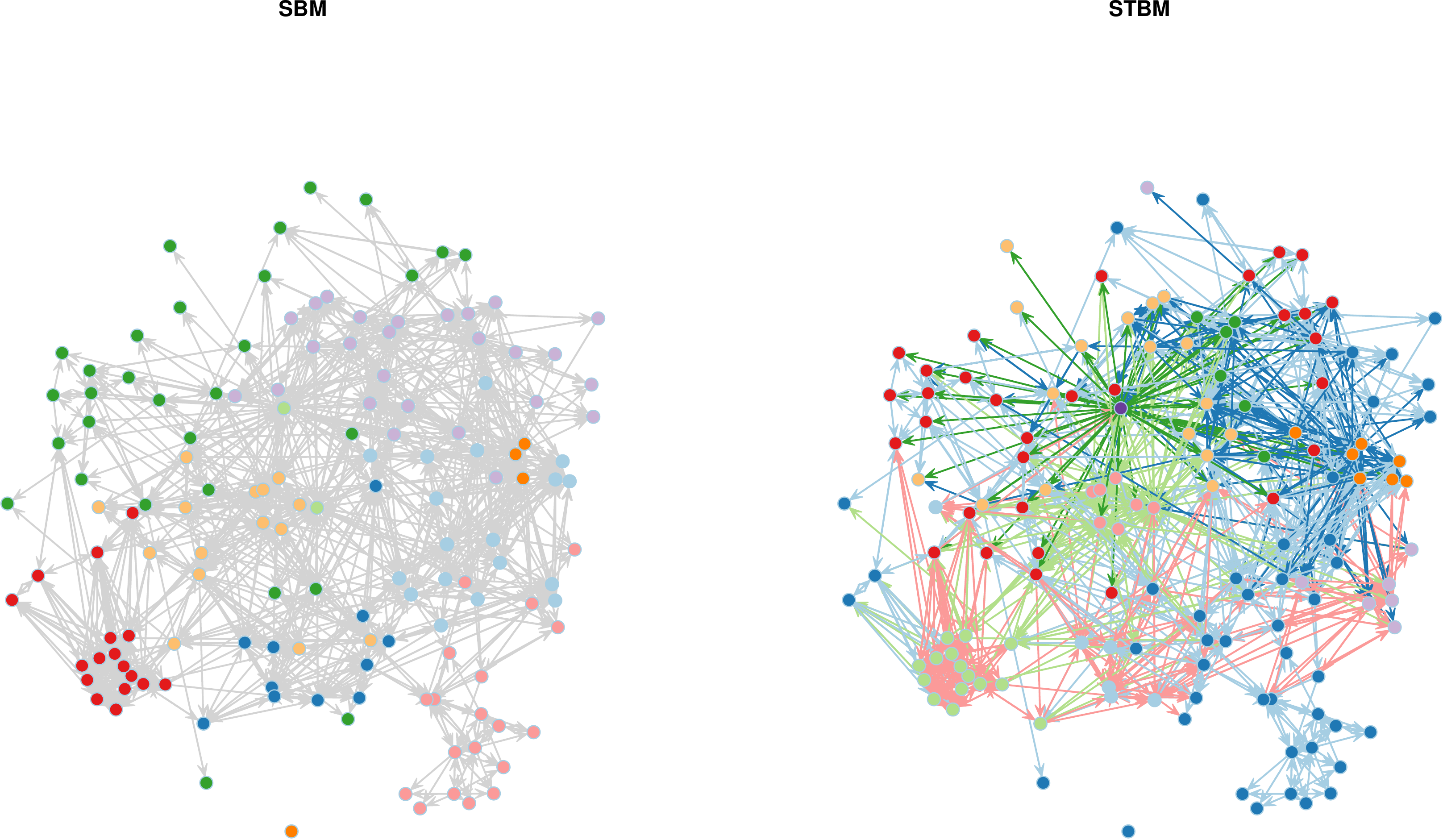}
\par\end{centering}

\protect\caption{\label{fig:Enron-SBMvsSTBM}Clustering results with SBM (left) and
STBM (right) on the Enron data~set. The selected number of groups
for SBM is $Q=8$ whereas STBM selects 10 groups and 5 topics.}
\end{figure}

Figure~\ref{fig:Enron-metaReso} presents a visual summary of connexion
probabilities between groups (the estimated $\pi$ matrix) and majority
topics for group interactions. A few elements deserve to be highlighted
in view of this summary. First, group 10 contains a single individual
who has a central place in the network and who mostly discusses about
logistic issues (topic 4) with groups 4, 5, 6 and 7. Second, group
8 is made of 6 individuals who mainly communicates about Enron activities
in Afghanistan (topic 2) between them and with other groups. Finally,
groups 4 and 6 seem to be more focused on trading activities (topic
1) whereas groups 1, 3 and 9 are dealing with technical issues on
gas deliveries (topic 5).

As a comparison, the network has also been processed with SBM, using
the mixer package~\citep{mixer}. The chosen number $K$ of groups
by SBM was 8. Figure~\ref{fig:Enron-SBMvsSTBM} allows to compare
the partitions of nodes provided by SBM and STBM. One can observe
that the two partitions differ on several points. On the one hand,
some clusters found by SBM (the bottom-left one for instance) have
been split by STBM since some nodes use different topics than the
rest of the community. On the other hand, SBM isolates two ``hubs''
which seem to have similar behaviors. Conversely, STBM identifies
a unique ``hub'' and the second node is gathered with other nodes,
using similar discussion topics. STBM has therefore allowed a better
and deeper understanding of the Enron network through the combination
of text contents with network structure.

\subsection{Analysis of the Nips co-authorship network}

This second network is a co-authorship network within a scientific
conference: the Neural Information Processing Systems (Nips) conference.
The conference was initially mainly focused on computational neurosciences
and is nowadays one of the famous conferences in statistical learning
and artificial intelligence. We here consider the data between the
1988 and 2003 editions (Nips 1--17). The data set, available at \url{http://robotics.stanford.edu/~gal/data.html},
contains the abstracts of 2~484 accepted papers from 2~740 contributing
authors. The vocabulary used in the paper abstracts has 14~036 words.
Once the co-authorship network reconstructed, we have an undirected
network between 2~740 authors with 22~640 textual edges.

\begin{figure}[p!]
\begin{centering}
\includegraphics[width=0.75\columnwidth]{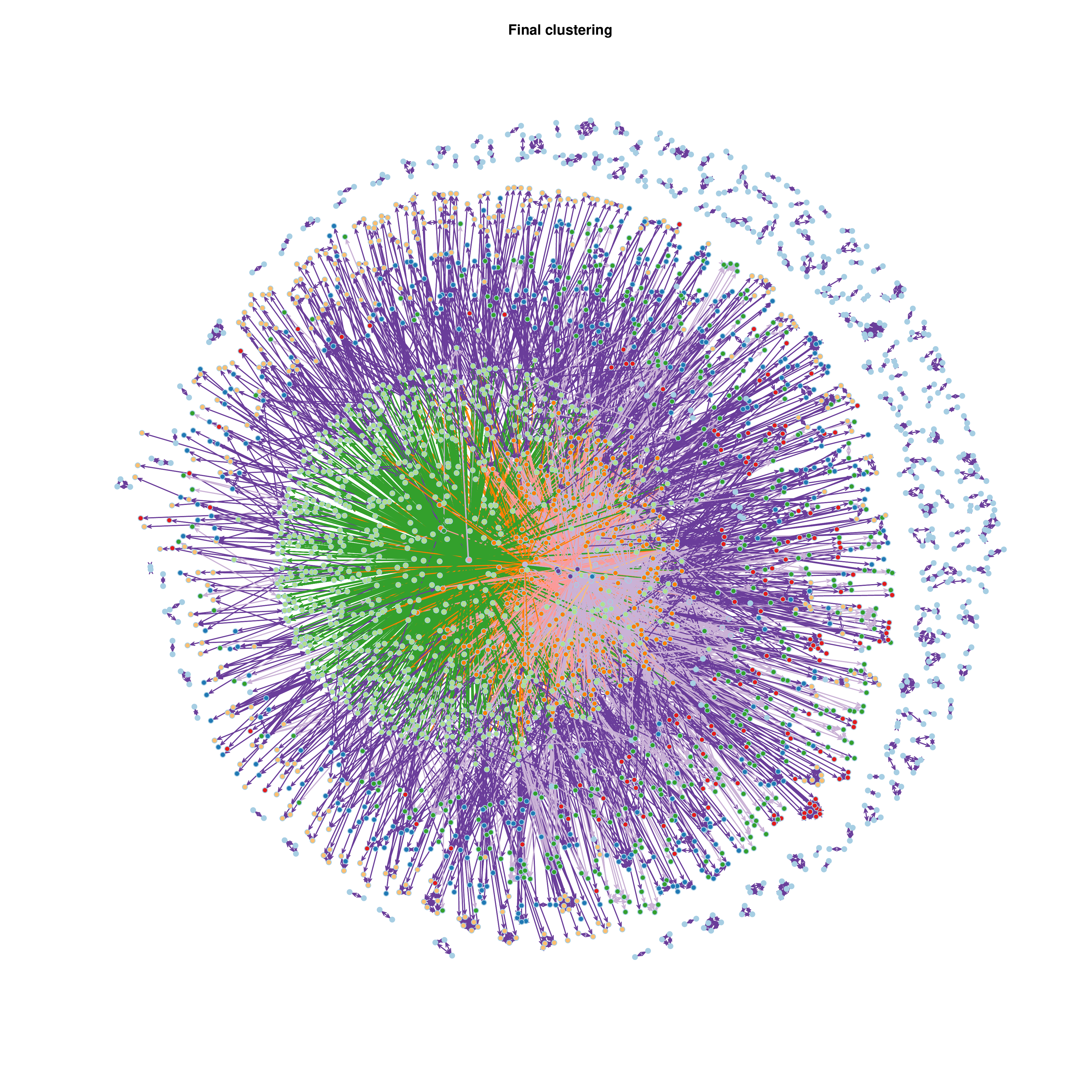}\includegraphics[width=0.25\columnwidth]{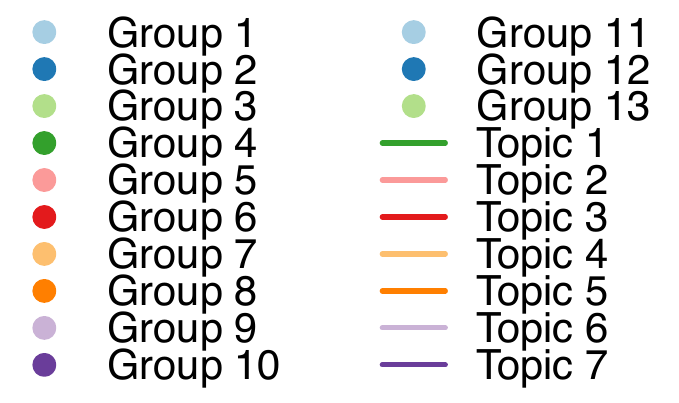}
\par\end{centering}

\protect\caption{\label{fig:Nips-Clustering-result}Clustering result with STBM on
the Nips co-authorship network.}
\end{figure}

\begin{figure}[p!]
\begin{centering}
\includegraphics[height=0.25\textheight]{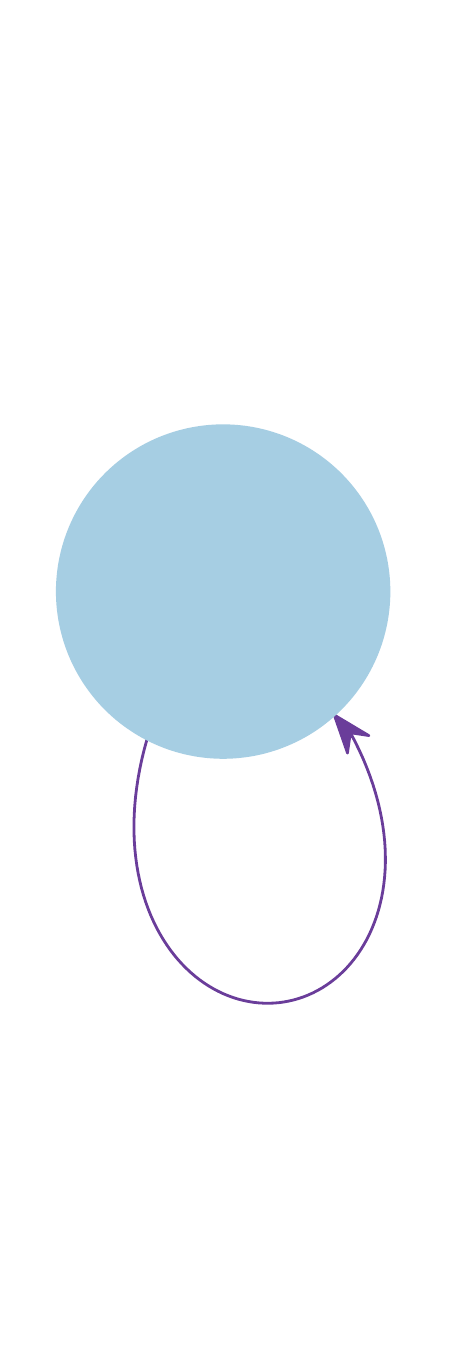}\includegraphics[angle=90,height=0.5\linewidth]{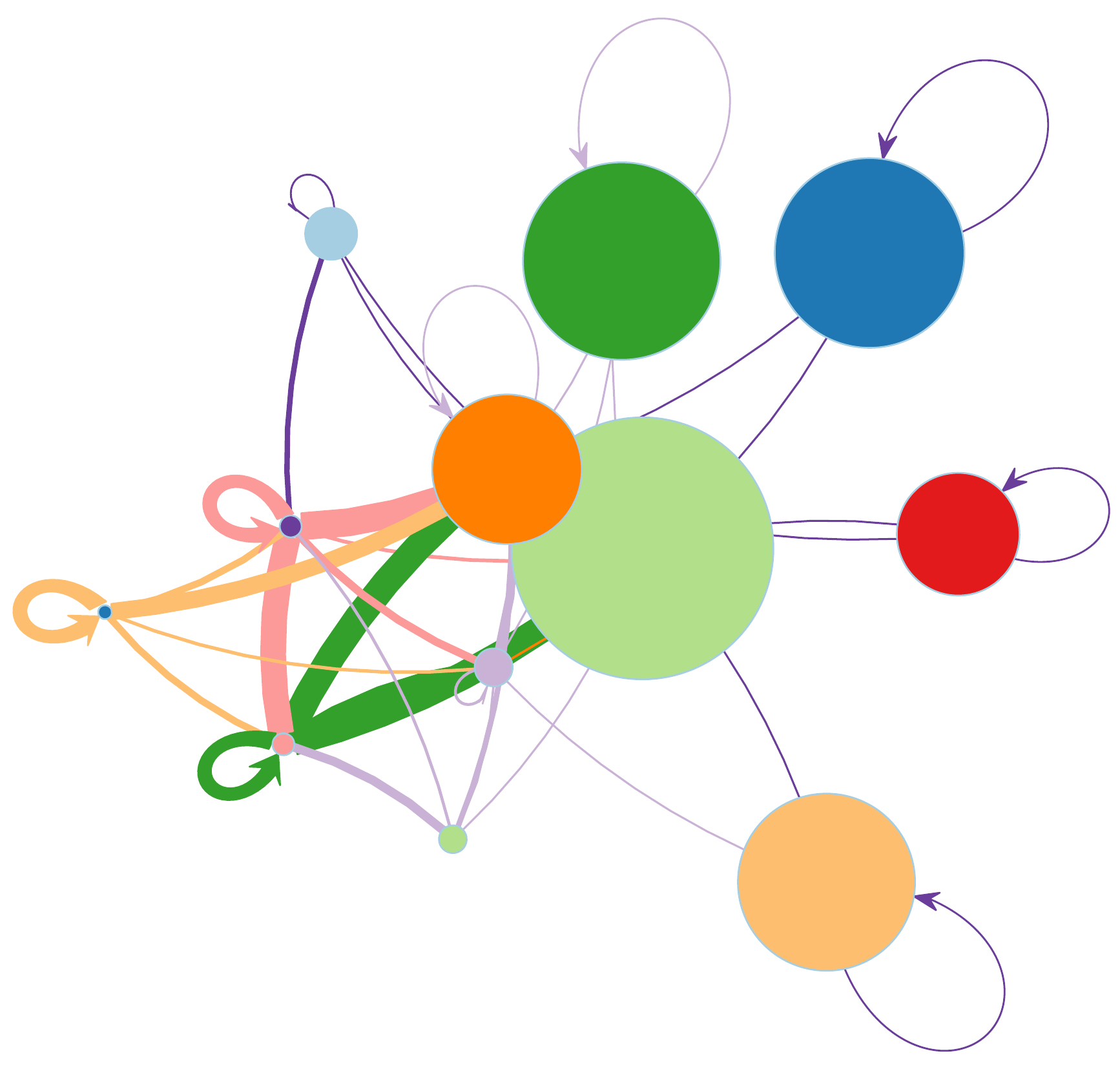}\includegraphics[width=0.25\columnwidth]{images/Nips-Legend.pdf}
\par\end{centering}

\protect\caption{\label{fig:Nips-metaReso}Nips co-authorship network: summary of connexion
probabilities between groups ($\pi$, edge widths), group proportions
($\rho$, node sizes) and most probable topics for group interactions
(edge colors).}
\end{figure}

\begin{figure}[t]
\begin{centering}
\includegraphics[width=0.85\columnwidth]{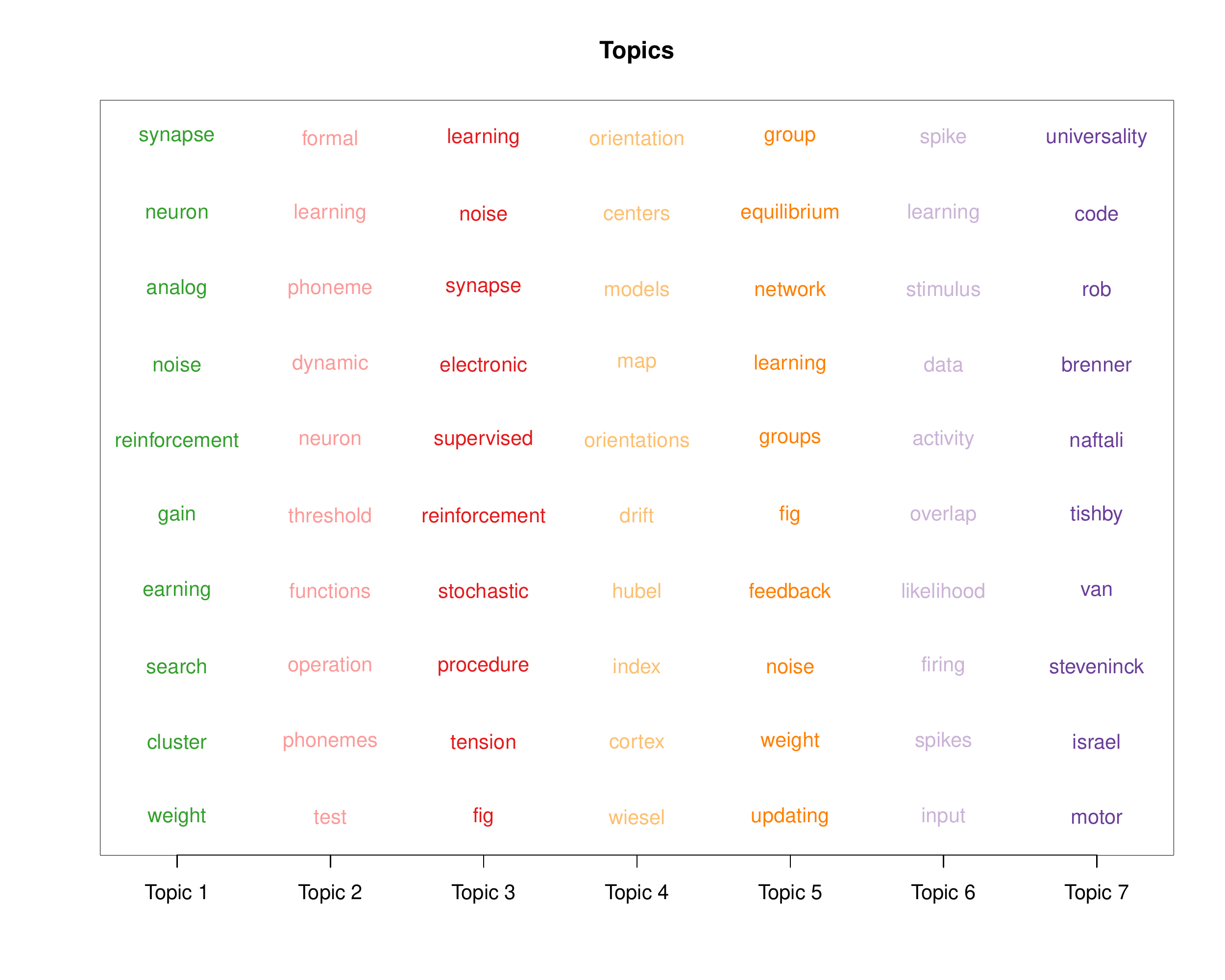}
\par\end{centering}

\protect\caption{\label{fig:Nips-Terms}Most specific words for the 5 found topics
with STBM on the Nips co-authorship network.}
\end{figure}

We applied STBM on this large data set and the selected model by ICL
was $(Q,K)=(13,7)$.  The values of ICL are presented in Figure~4 of the supplementary
  material. Note that the values of the criterion for $K > Q$ are not
  indicated since we found ICL to have higher values for $K \leq Q$ on
  this data set. It is worth noticing that STBM chose here a limited number of topics compared to what a simple LDA analysis of the data would have provided. Indeed, STBM looks for topics which are useful for clustering the nodes. In this sense, the topics of STBM may be slightly different than those of LDA. Figure~\ref{fig:Nips-Clustering-result} shows
the clustering obtained with STBM for $13$ groups of nodes and $7$
topics.  Due to size and density of the network, the visualization
and interpretation from this figure are actually tricky. Fortunately,
the meta-view of the network shown by Figure~\ref{fig:Nips-metaReso}
is of a greater help and allows to get a clear idea of the network
organization. To this end, it is necessary to first picture out
the meaning of the found topics (see Figure~\ref{fig:Nips-Terms}):
\begin{itemize}
\item Topic 1 seems to be focused on neural network theory, which was and
still is a central topic in Nips,
\item Topic 2 is concerned with phoneme classification or recognition,
\item Topic 3 is a more general topic about statistical learning and artificial
intelligence,
\item Topic 4 is about Neuroscience and focuses on experimental works about
the visual cortex,
\item Topic 5 deals with network learning theory, 
\item Topic 6 is also about Neuroscience but seems to be more focused on
EEG,
\item Topic 7 is finally devoted to neural coding, \emph{i.e.} characterizing
the relationship between the stimulus and the individual responses.
\end{itemize}
In light of these interpretations, we can eventually comment some
specific relationships between groups. First of all, we have an obvious
community (group 1) which is disconnected with the rest of the network
and which is focused on neural coding (topic 7). One can also clearly
identifies, on both Figure~\ref{fig:Nips-metaReso} and the reorganized
adjacency matrix (Figure~6 of the supplementary material)
that groups 2, 5 and 10 are three ``hubs'' of a few individuals.
Group~2 seems to mainly work on the visual cortex understanding whereas
group~10 is focused on phoneme analysis. Group~5 is mainly concerned
with the general neural network theory but has also collaborations
in phoneme analysis. From a more general point of view, topics~6
and~7 seem to be popular themes in the network. Notice that group~3 has a specific behavior in the network since people in this cluster publish preferentially with people in other groups than together. This is the exact definition of a disassortative cluster. This appears clearly on Figure~6 of the supplementary material.
It is also of interest to notice that statistical learning and artificial intelligence (which
are probably now 90\% of the submissions at Nips) were not yet by
this time proper thematics. They were probably used more as tools
in phoneme recognition studies and EEG analyses. This is confirmed by the fact that words used in topic 3 are less specific to the topic and are frequently used in other topics as well 
(see Figure~7 of the supplementary material).

As a conclusive remark on this network, STBM has proved its ability
to bring out concise and relevant analyses on the structure of a large
and dense network. In this view, the meta-network of Figure~\ref{fig:Nips-metaReso}
is a great help since it summarizes several model parameters of STBM.

\section{Conclusion}

This work has  introduced a probabilistic model,  named the stochastic
topic bloc model  (STBM), for the modeling and  clustering of vertices
in networks 
with textual  edges. The  proposed model allows  the modeling  of both
directed  and  undirected  networks, authorizing  its  application  to
networks   of    various   types   (communication,    social   medias,
co-authorship, ...). A classification variational EM (C-VEM) algorithm
has  been proposed  for model  inference and  model selection  is done
through  the ICL  criterion. Numerical  experiments on  simulated data
sets have proved the effectiveness of the proposed methodology.  Two real-world networks (a communication and a
co-authorship network) have also been studied using the STBM model and
insightful results have been exhibited. It is worth noticing that STBM
has been  applied to a  large co-authorship network with  thousands of
vertices, proving  the scalability of  our approach.

Further work  may include  the extension of  the STBM  model to
  dynamic  networks and  networks  with covariate  information on  the
  nodes and / or edges. The extension to the dynamic framework 
would be possible by adding for instance a state space model over group and topics
proportions. Such  an approach has  already been used with  success on
SBM-like  models,  such  as  in~\cite{Zreik2016}.  It  would  also  be
possible to take  into account covariate information  available on the
nodes   by  adopting   a  mixture   of  experts   approach,  such   as
in~\cite{gormley2010mixture}. Extending the  STBM  model to  overlapping
clusters  of  nodes  would  be  another natural  idea.  It  is  indeed
commonplace  in social  analysis  to allow  individuals  to belong  to
multiple groups (family, work, friends, ...). One possible choice would
be       to       derive       an       extension       of     the  MMSBM
model~\citep{airoldi2008mixed}. However, this would increase significantly the parameterization of the model. Finally, STBM could also be adapted in order to take into account the intensity or the type of communications between individuals.

\begin{acknowledgements}
The authors would like to greatly thank the editor and the two reviewers for their helpful remarks on the first version of this paper, and Laurent Bergé for his kind suggestions and the development  of visualization tools.
\end{acknowledgements}

\appendix
\section{Appendix}

\subsection{Optimization of $R(Z)$}\label{app:rz}

The   VEM  update   step  for   each  distribution   $R(Z_{ij}^{dn})$,
$A_{ij}=1$, is given by
\begin{equation}\label{eq:aoptZ}
\begin{aligned}
  \log        R(Z_{ij}^{dn})        &=        \mathrm{E}_{Z^{\backslash
      i,j,d,n},\theta}[\log p(W|A, Z, \beta) + \log p(Z|A, Y, \theta)]
  + \cst \\
  &=         \sum_{k=1}^{K}
  Z_{ij}^{dnk}\sum_{v=1}^{V}W_{ij}^{dnv}\log        \beta_{kv}       +
  \sum_{q,r}^{Q}Y_{iq}Y_{jr}\sum_{k=1}^{K}Z_{ij}^{dnk}\Esp_{\theta_{qr}}[\log
  \theta_{qrk}] + \cst \\
  &=
  \sum_{k=1}^{K}Z_{ij}^{dnk}\left(\sum_{v=1}^{V}W_{ij}^{dnv}\log
    \beta_{kv} + \sum_{q,r}^{Q}Y_{iq}Y_{jr}\Big(\psi(\gamma_{qrk}-\psi(\sum_{k=1}^{K}\gamma_{qrk})\Big)\right) + \cst,
\end{aligned}
\end{equation}
where all terms that do not depend on $Z_{ij}^{dn}$ have been put into
the constant term $\cst$.  Moreover, $\psi(\cdot)$ denotes the digamma
function. The functional form of a multinomial
distribution is then recognized in (\ref{eq:aoptZ})
\begin{equation*}
  R(Z_{ij}^{dn})=\mathcal{M}\left(Z_{ij}^{dn};1,\phi_{ij}^{dn}=(\phi_{ij}^{dn1},\dots, \phi_{ij}^{dnK})\right),
\end{equation*}
where 
\begin{equation*}
  \phi_{ij}^{dnk}     \propto     \left(\prod_{v=1}^{V}
  \beta_{kv}^{W_{ij}^{dnv}}\right)\prod_{q,r}^{Q}\exp\Big(\psi(\gamma_{qrk}-\psi(\sum_{l=1}^{K}\gamma_{qrl})\Big)^{Y_{iq}Y_{jr}}.
\end{equation*}
$\phi_{ij}^{dnk}$ is the (approximate) posterior distribution of
words $W_{ij}^{dn}$ being in topic $k$. 

\subsection{Optimization of $R(\theta)$}\label{app:rtheta}

The VEM update step for distribution $R(\theta)$ is given by
\begin{equation*}
  \begin{aligned}
  \log R(\theta) &= \Esp_{Z}[\log p(Z|A, Y, \theta)] + \cst \\
  &=                            \sum_{i                           \neq
    j}^{M}A_{ij}\sum_{d=1}^{D_{ij}}\sum_{n=1}^{N_{ij}^{d}}\sum_{q,r}^{Q}Y_{iq}Y_{jr}\sum_{k=1}^{K}\Esp_{Z_{ij}^{dn}}[Z_{ij}^{dnk}]\log
  \theta_{qrk}   +   \sum_{q,r}^{Q}\sum_{k=1}^{K}(\alpha_k  -   1)\log
  \theta_{qrk} + \cst \\
&=    \sum_{q,r}^{Q}\sum_{k=1}^{K}\left(\alpha_{k}    +    \sum_{i    \neq    j}^{M}
  A_{ij}Y_{iq}Y_{jr}\sum_{d=1}^{N_{ij}^{d}}\sum_{n=1}^{N_{ij}^{dn}}\phi_{ij}^{dnk}-1\right)\log \theta_{qrk}
+ \cst. 
  \end{aligned}
  \end{equation*}
We  recognize   the  functional  form   of  a  product   of  Dirichlet
distributions
\begin{equation*}
  \begin{aligned}
  R(\theta)                                                          =
  \prod_{q,r}^{Q}\mathrm{Dir}(\theta_{qr};\gamma_{qr}=(\gamma_{qr1},\dots, \gamma_{qrK})),
  \end{aligned}
\end{equation*}
where 
\begin{equation*}
  \gamma_{qrk} = \alpha_{k} +   \sum_{i    \neq    j}^{M}
  A_{ij}Y_{iq}Y_{jr}\sum_{d=1}^{N_{ij}^{d}}\sum_{n=1}^{N_{ij}^{dn}}\phi_{ij}^{dnk}.
\end{equation*}

\subsection{Derivation of the lower bound $\tilde{\mathcal{L}}\left(R(\cdot);
    Y, \beta\right)$}\label{app:bound}

The lower bound $\tilde{\mathcal{L}}\left(R(\cdot);
    Y, \beta\right) $ in (\ref{eq:tlowerBound}) is given by
  \begin{equation}\label{eq:aboundLDA}
    \begin{aligned}
      \tilde{\mathcal{L}}\left(R(\cdot);
    Y, \beta\right) &= \sum_{Z}\int_{\theta}R(Z,\theta)
\log \frac{p(W, Z, \theta|A, Y,\beta)}{R(Z,\theta)} d\theta \\
&= \Esp_{Z}[\log p(W|A, Z, \beta)] + \Esp_{Z, \theta}[\log p(Z|A, Y,
\theta)]  +  \Esp_{\theta}[\log  p(\theta)] -  \Esp_{Z}[\log  R(Z)]  -
\Esp_{\theta}[\log R(\theta)] \\ 
&=                             \sum_{i                            \neq
  j}^{M}A_{ij}\sum_{d=1}^{D_{ij}}\sum_{n=1}^{N_{ij}^{dn}}\sum_{k=1}^{K}\phi_{ij}^{dnk}\sum_{v=1}^{V}W_{ij}^{dnv}\log
\beta_{kv} \\
&\quad\quad + \sum_{i \neq
  j}^{M}A_{ij}\sum_{d=1}^{D_{ij}}\sum_{n=1}^{N_{ij}^{dn}}\sum_{q,r}^{Q}Y_{iq}Y_{jr}\sum_{k=1}^{K}\phi_{ij}^{dnk}\Big(\psi(\gamma_{qrk})-\psi(\sum_{l=1}^{K}\gamma_{qrl})\Big)
\\
&\quad\quad                 +                 \sum_{q,r}^{Q}\left(\log
  \Gamma(\sum_{l=1}^{K}\alpha_{k})                                 -
  \sum_{l=1}^{K}\log\Gamma(\alpha_{l})                             +
  \sum_{k=1}^{K}(\alpha_{k}-1)\Big(\psi(\gamma_{qrk})-\psi(\sum_{l=1}^{K}\gamma_{qrl})\Big)\right)
\\
&\quad\quad -    \sum_{i                            \neq
  j}^{M}A_{ij}\sum_{d=1}^{D_{ij}}\sum_{n=1}^{N_{ij}^{dn}}\sum_{k=1}^{K}\phi_{ij}^{dnk}\log
\phi_{ij}^{dnk} \\
&\quad\quad -   \sum_{q,r}^{Q}\left(\log
  \Gamma(\sum_{l=1}^{K}\gamma_{qrl})                                 -
  \sum_{l=1}^{K}\log\Gamma(\gamma_{qrl})                             +
  \sum_{k=1}^{K}(\gamma_{qrk}-1)\Big(\psi(\gamma_{qrk})-\psi(\sum_{l=1}^{K}\gamma_{qrl})\Big)\right)
    \end{aligned}
  \end{equation}

\subsection{Optimization of $\beta$}\label{app:beta}

In order to maximize the lower bound $\tilde{\mathcal{L}}\left(R(\cdot);
    Y, \beta\right) $, we isolate the terms in (\ref{eq:aboundLDA}) that depend on $\beta$ and
  add    Lagrange    multipliers    to   satisfy    the    constraints
  $\sum_{v=1}^{V}\beta_{kv}=1,\forall k$
  \begin{equation*}
    \tilde{\mathcal{L}}_{\beta} =   \sum_{i                            \neq
  j}^{M}A_{ij}\sum_{d=1}^{D_{ij}}\sum_{n=1}^{N_{ij}^{dn}}\sum_{k=1}^{K}\phi_{ij}^{dnk}\sum_{v=1}^{V}W_{ij}^{dnv}\log
\beta_{kv} + \sum_{k=1}^{K}\lambda_{k}(\sum_{v=1}^{V}\beta_{kv}-1).
  \end{equation*}
Setting the derivative, with respect to $\beta_{kv}$, to zero, we find
\begin{equation*}
  \beta_{kv}\propto \sum_{i \neq
  j}^{M}A_{ij}\sum_{d=1}^{D_{ij}}\sum_{n=1}^{N_{ij}^{dn}}\phi_{ij}^{dnk}W_{ij}^{dnv}.
\end{equation*}

\subsection{Optimization of $\rho$}\label{app:rho}

Only the distribution $p(Y|\rho)$ in  the complete data log-likelihood
$\log p(A,  Y|\rho, \pi)$  depends on the  parameter vector  $\rho$ of
cluster proportions. Taking  the log and adding  a Lagrange multiplier
to satisfy the constraint $\sum_{q=1}^{Q}\rho_{q}=1$, we have
\begin{equation*}
  \log p(Y|\rho) = \sum_{i=1}^{M}\sum_{q=1}^{Q}Y_{iq}\log \rho_{q}.
\end{equation*}
Taking the derivative with respect $\rho$ to zero, we find 
\begin{equation*}
  \rho_{q} \propto \sum_{i=1}^{M}Y_{iq}.
\end{equation*}

\subsection{Optimization of $\pi$}\label{app:pi}

Only the distribution $p(A|Y, \pi)$ in the complete data log-likelihood
$\log p(A,  Y|\rho, \pi)$  depends on the  parameter matrix $\pi$ of
connection probabilities.  Taking the  log we have
\begin{equation*}
  \log        p(A|Y,         \pi)        =         \sum_{i        \neq
    j}^{M}\sum_{q,r}^{Q}Y_{iq}Y_{jr}\Big(A_{ij}\log\pi_{qr}          +
  (1-A_{ij})\log(1-\pi_{qr})\Big).
\end{equation*}
Taking the derivative with respect to $\pi_{qr}$ to zero, we obtain
\begin{equation*}
  \pi_{qr} = \frac{   \sum_{i        \neq
    j}^{M}\sum_{q,r}^{Q}Y_{iq}Y_{jr}A_{ij}}{   \sum_{i        \neq
    j}^{M}\sum_{q,r}^{Q}Y_{iq}Y_{jr}}.
\end{equation*}

\subsection{Model selection}\label{app:ICL}

Assuming that the prior distribution over the model parameters $(\rho,
\pi, \beta)$ can be factorized, the integrated complete data log-likelihood $\log p(A, W, Y|K, Q)$ is given by
\begin{equation*}
\begin{aligned}
  \log  p(A, W,  Y|K, Q)  &=  \log \int_{\rho,\pi,\beta}  p(A, W,  Y, \rho,  \pi,
    \beta|K, Q)
    d\rho d\pi d\theta \\
  &=   \log    \int_{\rho,\pi,\beta}   p(A,    W,   Y|\rho,   \pi,    \beta,   K,
    Q)p(\rho|Q)p(\pi|Q)p(\beta|K)d\rho d\pi d\beta.
\end{aligned}
\end{equation*}
Note that the dependency on $K$ and  $Q$ is made explicit here, in all
expressions. In  all other sections of  the paper, we did  not include
these terms  to keep the  notations uncluttered. We find
\begin{equation}\label{eq:completeInte2}
  \begin{aligned}
    \log    p(A,   W,    Y|K,    Q)   &=    \log   \int_{\rho,    \pi,
      \beta}\left(\sum_{Z}\int_{\theta}p(A, W, Y, Z, \theta|\rho, \pi,
      \beta, K, Q)d\theta\right) p(\rho|Q)p(\pi|Q)p(\beta|K)d\rho d\pi
    d\beta \\
    &=  \log   \int_{\rho,    \pi,
      \beta}\left(\sum_{Z}\int_{\theta}p(W, Z, \theta|A, Y, 
      \beta, K, Q)p(A, Y|\rho, \pi, Q)d\theta\right) p(\rho|Q)p(\pi|Q)p(\beta|K)d\rho d\pi
    d\beta \\
    &= \log   \int_{\rho,    \pi,
      \beta}p(W|A, Y, 
      \beta, K, Q) p(A|Y, \pi, Q)p(Y|\rho, Q)p(\rho|Q)p(\pi|Q)p(\beta|K)d\rho d\pi
    d\beta \\
    &= \log \int_{\beta}p(W|A, Y, 
      \beta, K,  Q) p(\beta|K)  d\beta +  \log \int_{\pi}  p(A|Y, \pi,
      Q)p(\pi|Q)d\pi \\
&\quad\quad+ \log \int_{\rho}p(Y|\rho, Q)p(\rho|Q)d\rho.
  \end{aligned}
\end{equation}
Following  the derivation  of the  ICL criterion,  we apply  a Laplace
(BIC-like)   approximation   on   the    second   term   of   Equation
(\ref{eq:completeInte2}).  Moreover,  considering   a  Jeffreys  prior
distribution for $\rho$ and using Stirling formula for large values of
$M$, we obtain
\begin{equation*}
  \log \int_{\pi}  p(A|Y, \pi,
      Q)p(\pi|Q)d\pi \approx \max_{\pi}\log p(A|Y, \pi,
      Q) - \frac{Q^2}{2}\log M(M-1),
\end{equation*}
as well as 
\begin{equation*}
  \log \int_{\rho}p(Y|\rho, Q)p(\rho|Q)d\rho  \approx \max_{\rho} \log
  p(Y|\rho, Q) - \frac{Q-1}{2}\log M.
\end{equation*}
For         more         details,          we         refer         to
\cite{articlebiernacki2000}.  Furthermore,  we emphasize  that  adding
these two  approximations leads  to the ICL criterion  for the
SBM model, as derived by \cite{daudin2008mixture} 
\begin{equation*}
\begin{aligned}
  ICL_{SBM} &=  \max_{\pi}\log p(A|Y, \pi,
      Q) - \frac{Q^2}{2}\log M(M-1) + \max_{\rho} \log
  p(Y|\rho, Q) - \frac{Q-1}{2}\log M \\
  &=  \max_{\rho, \pi}  \log p(A,Y|\rho,  \pi, Q)  - \frac{Q^2}{2}\log
  M(M-1) - \frac{Q-1}{2}\log M.
\end{aligned}
\end{equation*}
In \cite{daudin2008mixture},  $M(M-1)$ is  replaced by  $M(M-1)/2$ and
$Q^2$ by $Q(Q+1)/2$ since they considered undirected networks.

Now, it is  worth taking a closer  look at the first  term of Equation
(\ref{eq:completeInte2}).  This term  involves a  marginalization over
$\beta$. Let us emphasize that $p(W|A, Y, \beta, K, Q)$ is related to the LDA model and involves a
marginalization over $\theta$ (and $Z$). Because we aim at approximating
the first term of Equation
(\ref{eq:completeInte2}), also with a Laplace (BIC-like) approximation, it is crucial
to identify  the number of  observations in the  associated likelihood
term $p(W|A, Y, \beta, K, Q)$. As
pointed  out  in  Section  \ref{ssection:probModel},  given  $Y$  (and
$\theta$),  it is  possible to  reorganize the  documents in  $W$ as
$W=(\tilde{W}_{qr})_{qr}$   is  such   a   way  that   all  words   in
$\tilde{W}_{qr}$   follow   the   same   mixture   distribution   over
topics. Each  aggregated document $\tilde{W}_{qr}$ has  its own vector
$\theta_{qr}$  of topic  proportions and  since the  distribution over
$\theta$  factorizes  ($p(\theta)=\prod_{q,r}^{Q}p(\theta_{qr}))$,  we
find
\begin{equation*}
\begin{aligned}
  p(W|A, Y, \beta, K, Q) &=  \int_{\theta} p(W |A, Y, \theta, \beta, K,
  Q)p(\theta|K, Q)d\theta \\
  &= \prod_{q,r}^{Q}\int_{\theta_{qr}}p(\tilde{W}_{qr}|\theta_{qr},
  \beta, K, Q)p(\theta_{qr}| K)d\theta_{qr} \\
  &= \prod_{q,r}^{Q} \ell (\tilde{W}_{qr}|\beta, K, Q),
\end{aligned}
\end{equation*}
where $\ell  (\tilde{W}_{qr}|\beta, K,  Q)$ is exactly  the likelihood
term of  the LDA model  associated with document  $\tilde{W}_{qr}$, as
described in \cite{LDA}. Thus
\begin{equation}\label{eq:inteLDA}
    \log \int_{\beta}p(W|A, Y, 
      \beta, K, Q) p(\beta|K) d\beta = \log \int_{\beta} p(\beta|K) \prod_{q,r}^{Q} \ell (\tilde{W}_{qr}|\beta, K, Q)d\beta.
\end{equation}
Applying  a Laplace  approximation on  Equation (\ref{eq:inteLDA})  is
then equivalent  to deriving  a BIC-like criterion  for the  LDA model
with documents in $W=(\tilde{W}_{qr})_{qr}$.  In the LDA model, the number of
observations  in  the  penalization  term  of BIC  is  the  number  of
documents \citep[see][for instance]{than2012}. In our case, this leads to
\begin{equation}
   \log \int_{\beta}p(W|A, Y, 
      \beta, K, Q) p(\beta|K) d\beta \approx \max_{\beta} \log p(W|A, Y, 
      \beta, K, Q) - \frac{K(V-1)}{2}\log Q^2.
\end{equation}
Unfortunately, $\log p(W|A, Y, 
      \beta, K, Q)$  is not tractable and so we  propose to replace it
      with its variational  approximation $\tilde{\mathcal{L}}$, after
      convergence of the C-VEM algorithm. By analogy with $ICL_{SBM}$,
      we call the corresponding criterion $BIC_{LDA|Y}$ such that
      \begin{equation*}
        \log p(A, W, Y|K, Q) \approx BIC_{LDA|Y} + ICL_{SBM}.
      \end{equation*}

\end{document}